\documentclass[prd,preprint,superscriptaddress,amsmath,amssymb,nofootinbib]{revtex4-1}
\usepackage{graphicx,color,slashed}

\usepackage{epstopdf}
\usepackage{slashed}
\usepackage{multirow}

\usepackage[bookmarks=true,bookmarksnumbered=true,bookmarkstype=toc, colorlinks=true, 
citecolor=blue, linkcolor=blue]{hyperref} 
\usepackage{units}

\newcommand{\lsim}{\raise0.3ex\hbox{$\;<$\kern-0.75em\raise-1.1ex\hbox{$\sim\;$}}}
\newcommand{\gsim}{\raise0.3ex\hbox{$\;>$\kern-0.75em\raise-1.1ex\hbox{$\sim\;$}}}

\begin{document}

{\begin{flushright}{EPHOU-19-018, KIAS-P19068, UME-PP-011}
\end{flushright}}

\title{Type II seesaw models with modular $A_4$ symmetry }

\author{Tatsuo Kobayashi}
\email{kobayashi@particle.sci.hokudai.ac.jp}
\affiliation{Department of Physics, Hokkaido University, Sapporo 060-0810, Japan}

\author{Takaaki Nomura}
\email{nomura@kias.re.kr}
\affiliation{School of Physics, KIAS, Seoul 02455, Korea}

\author{Takashi Shimomura}
\email{shimomura@cc.miyazaki-u.ac.jp}
\affiliation{Faculty of Education, Miyazaki University, Miyazaki, 889-2192, Japan}

\date{\today}

\begin{abstract}
We discuss type-II seesaw models adopting modular $A_4$ symmetry in supersymmetric framework. 
In our approach, the models are classified by the assignment of $A_4$ representations and modular weights for 
leptons and triplet Higgs fields.
Then neutrino mass matrix is characterized by modulus $\tau$ and two free parameters.
Carrying out numerical analysis, we find allowed parameter sets which can fit the neutrino oscillation data.
For the allowed parameter sets, we obtain the predictions in the neutrino sector such as CP violating phases and the lightest neutrino mass.
Finally we also show the predictions for the branching ratios of a doubly charged scalar boson, focusing on the case where the doubly charged scalar boson dominantly decays into charged leptons.
\end{abstract}
 \maketitle
\newpage

\section{Introduction}

Understanding of the flavor structure of leptons and quarks is one of the well motivated issues to construct a model 
of new physics beyond the standard model (SM).
In describing new physics, a new symmetry can play an important role to organize flavor structure.

The modular symmetry is a geometrical symmetry of torus and orbifold compactification, and very interesting, 
because it includes finite subgroups such as $S_3$, $A_4$, $S_4$, and $A_5$ \cite{deAdelhartToorop:2011re}.
Zero-modes in superstring theory on such compactifications and its low-energy  effective field theory 
transform non-trivially each other 
\cite{Lauer:1989ax,Lerche:1989cs,Ferrara:1989qb,Cremades:2004wa,Kobayashi:2017dyu,Kobayashi:2018rad}.\footnote{
See also Refs.~\cite{Kobayashi:2018bff,Baur:2019kwi,Kariyazono:2019ehj}.}
Inspired by these aspects, the framework of modular flavor symmetries have been recently proposed by~\cite{Feruglio:2017spp}
to realize more predictable structure in the quark and lepton sectors where a coupling can be transformed under a non-trivial representation of a non-Abelian discrete group.  
The typical groups in the framework are found on basis of  the $A_4$ modular group \cite{Feruglio:2017spp, Criado:2018thu, Kobayashi:2018scp, Okada:2018yrn, Nomura:2019jxj, Okada:2019uoy, deAnda:2018ecu, Novichkov:2018yse, Nomura:2019yft, Okada:2019mjf,Ding:2019zxk,Nomura:2019lnr,Kobayashi:2019xvz,Asaka:2019vev,Zhang:2019ngf}, $S_3$ \cite{Kobayashi:2018vbk, Kobayashi:2018wkl, Kobayashi:2019rzp, Okada:2019xqk}, $S_4$ \cite{Penedo:2018nmg, Novichkov:2018ovf, Kobayashi:2019mna,King:2019vhv,Okada:2019lzv,Criado:2019tzk,Wang:2019ovr}, $A_5$ \cite{Novichkov:2018nkm, Ding:2019xna,Criado:2019tzk}, multiple modular symmetries~\cite{deMedeirosVarzielas:2019cyj}, and double covering of $A_4$~\cite{Liu:2019khw} in which  masses, mixing angles, and CP phases for quarks and leptons are predicted.~\footnote{Several reviews are helpful to understand the non-Abelian group and its applications to flavor structures~\cite{Altarelli:2010gt, Ishimori:2010au, Ishimori:2012zz, Hernandez:2012ra, King:2013eh, King:2014nza, King:2017guk, Petcov:2017ggy}.}
Possible corrections coming from the K\"ahler potential are also considered in Ref.~\cite{Chen:2019ewa}, and
 a systematic approach to understand the origin of CP transformations is discussed in Ref.~\cite{Baur:2019kwi,Novichkov:2019sqv}.
Also CP violation in models with modular symmetry is discussed in Ref.~\cite{Kobayashi:2019uyt}.
In applying a modular symmetry it is especially interesting to consider a new physics model generating neutrino masses since 
we would obtain the predictions for signals of new physics and observables in the neutrino sector, which can be correlated each 
other.

In realizing small neutrino masses, the so-called type-II seesaw mechanism is one of the interesting ideas in which an $SU(2)$ triplet Higgs field is introduced~\cite{Magg:1980ut, Konetschny:1977bn, Hirsch:2008gh}.
The neutrino masses are generated through Yukawa interactions among the triplet and lepton doublets after the triplet developed  a vacuum expectation value (VEV). 
In this scenario, we have a doubly charged scalar boson from the triplet which couples to charged leptons.
The doubly charged scalar boson dominantly decays into the same sign charged lepton pair when the triplet VEV is less than around $10^{-4}$ GeV, 
and it can give clear signals at the collider experiments such as the LHC.
Importantly the branching ratios (BRs) of such decays are given by Yukawa couplings associated with neutrino mass generation and 
we can obtain some correlations among the BRs and neutrino parameters.

In this paper, we apply the modular $A_4$ symmetry to the type-II seesaw mechanism in the supersymmetric framework.
Then some possible models are classified by the assignments of $A_4$ representations and modular weights to the leptons and the Higgs triplet.
We then scan free parameters in these models and search for the region in which the neutrino oscillation data can be fitted.
For the allowed parameter sets, we show the predictions of observables in the neutrino sector.
Finally we show our predictions for the branching ratios of the doubly charged scalar boson applying the allowed parameter sets.

The paper is organized as follows.
In section 2,  we introduce our models.
In section 3, we perform a parameter scan to fit neutrino oscillation data and provide some predictions in observables in 
the neutrino sector.
Also, we show the branching ratios of the doubly charged scalar boson applying the parameter sets 
accommodating the neutrino oscillation data. 
Section 4 is our conclusion and discussions.
In Appendix \ref{apdx:multiplication-rule}, we give generators and multiplication rules used in this paper, and 
in Appendix \ref{apdx:lepton-mass}, we summarize formulae to fix the coupling coefficients for the Yukawa interactions associated with charged lepton masses.


\section{Models}

\begin{center} 
\begin{table}[tb]
\begin{tabular}{|c||c|c||c|c|c|c||}\hline\hline  
&\multicolumn{2}{c||}{ Lepton} & \multicolumn{4}{c||}{Higgs}   \\ \hline
  & ~$L$~ & ~$(e_{R},\mu_{R},\tau_{R})$ ~ & ~$T_1$~ & ~$T_2$~ & ~$H_u$~  & ~$H_d$~ 
  \\\hline 
 $SU(2)_L$ & ${\bf 2}$   & $1$  & ${\bf 3}$ & ${\bf 3}$ & ${\bf 2}$ & ${\bf 2}$    \\\hline 
$U(1)_Y$ & $-\frac12$  & $1$ & $1$ & $-1$  & $\frac12$  & $-\frac12$   \\\hline
 $A_4$ & $3$ & ${1,1'',1'}$ &Model (1), (3): $1^{\ }$  &(1), (3): $1^{\ }$&(1), (3): $1$ & $1$  \\
  &   &  &Model (2), (4): $1''$   &(2), (4): $1'$  &(2), (4): $1'$ &  \\ \hline
 $k_I$ & Model (1), (2): $-1$ \ & $-1$ & $0$ & $0$ & $0$ & $0$    \\
           & Model (3), (4): $-2$ \ &  $0[-2]$ for case A[B]  & &  & &   \\ \hline
\end{tabular}
\caption{Assignments under $SU(2)_L\times U(1)_Y\times A_4$ for lepton and scalar superfields. }
\label{tab:assigment1}
\end{table}
\end{center}

In this section we show type-II seesaw models with modular $A_4$ symmetry in the supersymmetric framework under which 
superfields of leptons are non-trivially transformed under the modular symmetry. 
In the type-II seesaw mechanism, we introduce two $SU(2)$ triplet superfields $T_1$ and $T_2$ which 
have hypercharges  $Y=1$ and $-1$, respectively; 
here we need two triplet superfields for gauge anomaly cancellation.
We then obtain a superpotential of the form
 \begin{equation}
w_\nu = {\bf y}_T L T_1 L +  \lambda_1 H_d T_1 H_d + \lambda_2 H_u T_2 H_u + M_T T_1 T_2,
\end{equation}
where $L$ is the superfield for the lepton doublet, and $H_u$ and $H_d$ are superfields for the Higgs doublets with hypercharges $\frac12$ and $-\frac12$, respectively. 
In the following discussion, we use the same symbols for the SM leptons and scalars as their superfields.
As in the minimal supersymmetric SM, they develop VEVs $\langle H_{u,d} \rangle = v_{u,d}/\sqrt{2}$ inducing SM fermion mass terms.
From the superpotential, we obtain the VEV of the neutral component of the $T_1$ scalar, denoted 
by $\langle T_1 \rangle = v_{T_1}$, as follows
\begin{align}
v_{T_1} =  {\lambda}_2 \frac{v_u^2}{2 M_T}.
\end{align}
The VEV provides neutrino mass term~\cite{Hirsch:2008gh} as we show below. 
The superpotential terms relevant to the charged lepton masses are written by
\begin{equation}
w_e = {\bf y}_e e_R H_d L +  {\bf y}_{\mu} \mu_R H_d L +  {\bf y}_\tau \tau_R H_d L,
\end{equation}
where superfields $\{e_R, \mu_R, \tau_R\}$ correspond to right-handed charged leptons.
These superpotential terms are required to be invariant under $A_4$ symmetry with vanishing modular weight. 
Here, the couplings can be modular forms associated with non-trivial $A_4$ representations and 
having non-zero modular weights.
More specifically, modular forms $f(\tau)$ are transformed as
\begin{align}
& f_i(\gamma\tau)= (c\tau+d)^k \rho(\gamma)_{ij} f_j(\tau)~, \\
& \left( \tau \longrightarrow \gamma\tau= \frac{a\tau + b}{c \tau + d}\ ,~~ {\rm where}~~ a,b,c,d \in \mathbb{Z}~~ {\rm and }~~ ad-bc=1, ~~ {\rm Im} [\tau]>0 ~ , \right), \nonumber 
\end{align}
where $\tau$ is modulus, $k$ is the modular weight and $\rho(\gamma)_{ij}$ indicate a unitary transformation matrix under $A_4$.
Similarly, a multiplet of chiral superfields transform
\begin{align}
& \phi^{(I)}\to(c\tau+d)^{k_I}\rho^{(I)}(\gamma)\phi^{(I)}.
\end{align}
Then models are distinguished by the 
assignments of $A_4$ representations and modular weights for the lepton, Higgs doublet and triplet superfields.
In Table~\ref{tab:assigment1}, we summarize the assignment of $A_4$ representations and modular weights to 
the superfields in our models.
With these representations and weights of the fields, those of the Yukawa couplings are fixed. Then, the structure of  the 
superpotential is determined. 
The other sectors are assumed to be the same as those of  the supersymmetric type II seesaw model~\cite{Hirsch:2008gh}, 
which we do not discuss in this paper~\footnote{ For models (2) and (4), we assign $1''$ representation to superfield associated with $u_R$ to obtain $Q_L H_u u_R$ term, and we need to introduce singlet scalar $\varphi$ under $1''$ with non-zero VEV to generate $H_u H_d$ term.}.

The Yukawa coupling constants can have the modular weights under the modular symmetry. 
The modular form of $A_4$ triplet with weight $2$, ${\bf Y}_{\bf 3}^{(2)}(\tau)$, is given by
\begin{equation}
{\bf Y}_{\bf 3}^{(2)}(\tau)=\begin{pmatrix}Y_1(\tau)\\Y_2(\tau)\\Y_3(\tau)\end{pmatrix}=
\begin{pmatrix}
1+12q+36q^2+12q^3+\dots \\
-6q^{1/3}(1+7q+8q^2+\dots) \\
-18q^{2/3}(1+2q+5q^2+\dots)\end{pmatrix}, \quad  q = e^{2 \pi i \tau},
\label{eq:triplet-yukawa}
\end{equation}
where $\tau$ is a complex number.
More precisely, the above modular forms can be written in terms of the Dedekind eta-function $\eta(\tau)$ and its derivative:
\begin{eqnarray} 
\label{eq:Y-A4}
Y_{1}(\tau) &=& \frac{i}{2\pi}\left( \frac{\eta'(\tau/3)}{\eta(\tau/3)}  +\frac{\eta'((\tau +1)/3)}{\eta((\tau+1)/3)}  
+\frac{\eta'((\tau +2)/3)}{\eta((\tau+2)/3)} - \frac{27\eta'(3\tau)}{\eta(3\tau)}  \right), \nonumber \\
Y_{2}(\tau) &=& \frac{-i}{\pi}\left( \frac{\eta'(\tau/3)}{\eta(\tau/3)}  +\omega^2\frac{\eta'((\tau +1)/3)}{\eta((\tau+1)/3)}  
+\omega \frac{\eta'((\tau +2)/3)}{\eta((\tau+2)/3)}  \right) ,  \\ 
Y_{3}(\tau) &=& \frac{-i}{\pi}\left( \frac{\eta'(\tau/3)}{\eta(\tau/3)}  +\omega\frac{\eta'((\tau +1)/3)}{\eta((\tau+1)/3)}  
+\omega^2 \frac{\eta'((\tau +2)/3)}{\eta((\tau+2)/3)}  \right)\,,
\nonumber
\end{eqnarray}
where $\omega = e^{2 \pi i/3}$. Equation \eqref{eq:triplet-yukawa} is their $q$-expansions.
The modular forms with higher weights can be constructed by products of ${\bf Y}_{\bf 3}^{(2)}(\tau)$. 
Singlet modular forms with weight $4$, $Y_{1}^{(4)}$ and $Y_{1'}^{(4)}$are given by
\begin{align}
Y_1^{(4)} = Y_1^2 + 2 Y_2 Y_3, \\
Y_{1'}^{(4)} = Y_3^2 + 2 Y_1 Y_2, 
\end{align}
where the modular form of the $1''$ representation with weight $4$ does not exist due to the relation $Y_2^2 + 2 Y_1 Y_3 =0$. 
Furthermore the triplet modular form with weight $4$, ${\bf Y}_{\bf 3}^{(4)}$, is constructed as 
\begin{equation}
 {\bf Y^{(4)}_{3}} \equiv
 \begin{pmatrix}
Y_{3,1}^{(4)}  \\
Y_{3,2}^{(4)} \\
Y_{3,3}^{(4)} \\
\end{pmatrix} 
 =
\begin{pmatrix}
Y^2_1 - Y_2 Y_3  \\
Y_3^2 - Y_1 Y_2 \\
Y_2^2 - Y_1 Y_3 \\
\end{pmatrix} \ .
\end{equation}

After the electroweak symmetry breaking, we obtain mass terms for leptons from the superpotentials 
$w_e$ and $w_\nu$ such that
\begin{align}
& \bar \ell_R M_E  \ell_L + h.c. = {\bf y}_e e_R \langle H_d \rangle L +  {\bf y}_\mu \mu_R \langle H_d \rangle L  +  {\bf y}_\tau \tau_R  \langle H_d \rangle L + h.c., \\ 
& \frac{1}{2}\bar \nu_L^c M_\nu  \nu_L =  \bar L^c {\bf y}_T \langle T_1 \rangle L_L,
\end{align}
where the flavor index is omitted.
The structure of the mass matrices is determined by assignments of modular $A_4$ representations.
In the following, we discuss them in each model.

\subsection{Model (1)}

In this model we can write the superpotential terms relevant to the neutrino masses as
\begin{equation}
\label{eq:SPnu1}
w_\nu = y {\bf Y}_{\bf 3}^{(2)} L T_1 L + \lambda_1 H_d T_1 H_d + \lambda_2 H_u T_2 H_u + M_T T_1 T_2,
\end{equation}
and the superpotential terms relevant to the charged lepton masses,
\begin{equation}
\label{eq:SPcharged}
w_e = \alpha e_R H_d \left(L {\bf Y}_{\bf 3}^{(2)} \right) + \beta \mu_R H_d \left( L {\bf Y}_{\bf 3}^{(2)} \right) + \gamma \tau_R H_d \left( L {\bf Y}_{\bf 3}^{(2)} \right).
\end{equation} 
The mass matrix for the charged leptons is given by 
\begin{align}
 L_{M_E} = \bar \ell_R M_E \ell_L, \quad 
 M_E = \tilde \gamma Y_3 \text{ diag}[\hat \alpha, \hat \beta, 1] 
\begin{pmatrix} \hat Y_1 & 1 & \hat Y_2 \\ \hat Y_2 & \hat Y_1 & \hat 1 \\ \hat 1 & \hat Y_2 & \hat Y_1 \end{pmatrix}, 
\label{eq:ME1}
\end{align}
where $\ell$ denotes three generations of charged leptons, $\tilde \gamma \equiv v_d \gamma/\sqrt{2}$, $\hat \alpha \equiv \alpha/\gamma$, $\hat \beta = \beta/\gamma$ and $\hat Y_{1,2} \equiv Y_{1,2}/Y_3$. 
To obtain Eq.~\eqref{eq:ME1},  we used the multiplication rules given in Appendix \ref{apdx:multiplication-rule}.
As in the SM, we can diagonalize the mass matrix by transforming lepton fields, $\ell_{L(R)} \to V^e_{L(R)} \ell_{L(R)}$, providing 
${\rm diag}(m_e, m_\mu, m_\tau) = (V^e_R)^\dagger M_e V^e_L$. 
The parameters $\hat \alpha$ and $\hat \beta$ are determined to provide charged lepton mass eigenvalues as given in Appendix~\ref{apdx:lepton-mass}.

After the neutral component of $T_1$ developing its VEV, $v_{T_1}$, we obtain Majorana neutrino mass terms such as
\begin{align}
L_{M_\nu} &= \frac{y v_{T_1} }{3}  \bar \nu'^c_{L_i} \begin{pmatrix} 2 Y_1 & - Y_3 & - Y_2 \\ - Y_3 & 2 Y_2 & - Y_1 \\ -Y_2 & -Y_1 & 2 Y_3 \end{pmatrix}_{ij} \nu'_{L_j},
\end{align}
where $\nu'_{L_{i=1,2,3}}$ denotes the neutral fermion component of $L$. 
Note that $\nu'_{L_i}$s are not identified with $\nu_{e,\mu,\tau}$, the partners of the charged leptons in weak interaction, 
since they are in the basis where the charged lepton mass matrix is not diagonalized.
Then we find the lepton flavor basis by
\begin{equation}
(\nu'_1, \nu'_2, \nu'_3)^T = V^e_L(\nu_e, \nu_\mu, \nu_\tau)^T.
\end{equation}
Thus the neutrino mass matrix in the flavor basis is given by
\begin{equation}
m_\nu = \frac{2 y v_{T_1} }{3} (V^e_L)^T
\begin{pmatrix} 2 Y_1 & - Y_3 & - Y_2 \\ - Y_3 & 2 Y_2 & - Y_1 \\ -Y_2 & -Y_1 & 2 Y_3 \end{pmatrix} V^e_L.
\end{equation}
Notice that the mixing matrix $V^e_L$ is involved in the neutrino mass matrix.

\subsection{Model (2)}

In this model, we take the $A_4$ representations of $T_1,~T_2$ and $H_u$ as $1'',~1'$ and $1'$ while the other setting is the same as model (1).
Then the superpotential terms relevant to the neutrino masses are
\begin{equation}
w_\nu = y {\bf Y}^{(2)}_{\bf 3}(\tau) L T_1 L  + \lambda_2 H_u T_2 H_u + M_T T_1 T_2.
\end{equation}
Note that we do not have the $\lambda_1 H_d T_1 H_d$ term compared to Eq.~(\ref{eq:SPnu1}) where the term is irrelevant in realizing the type-II seesaw mechanism and absence of the term does not affect our analysis.
For the charged lepton mass term, the superpotential is the same as Eq.~(\ref{eq:SPcharged}).

The neutrino mass matrix in this case is
\begin{equation}
m_\nu = \frac{2 y v_{T_1} }{3} (V^e_L)^T
\begin{pmatrix} 2 Y_2 & - Y_1 & - Y_3 \\ - Y_1 & 2 Y_3 & - Y_2 \\ -Y_3 & -Y_2 & 2 Y_1 \end{pmatrix} V^e_L,
\end{equation}
where the structure is different from model (1).

\subsection{Model (3)}

In this model, we take the modular weight $-2$ for leptons and the assignment under the $A_4$ representation is the same as model (1).
Then the superpotential terms relevant to the neutrino masses are 
\begin{align}
w_\nu = & \ y_1 {\bf Y}_{\bf 3}^{(4)}(\tau) (L T_1 L)_{3} + y_2 Y_{1}^{(4)}(\tau) (L T_1 L)_{1} + y_3 Y_{1'}^{(4)}(\tau) (L T_1 L)_{1''} \\
& + \lambda_1 H_d T_1 H_d + \lambda_2 H_u T_2 H_u + M_T T_1 T_2.
\end{align}
In this case, we have additional terms with free parameters since $A_4$ singlet modular forms are also available when couplings should have the modular weight $4$.

For the charged lepton mass term, 
we consider two cases depending on the modular weight assignment for right-handed charged leptons. In cases A and B,  
the modular weights of $\ell_R$ are assigned to $0$ and $-2$, respectively.
Then case A has the same superpotential as Eq.~(\ref{eq:SPcharged}).
On the other hand, for case B we obtain the corresponding superpotential as 
\begin{equation}
\label{eq:SPcharged2}
w_e = \alpha e_R H_d \left(L {\bf Y}_{\bf 3}^{(4)} \right) + \beta \mu_R H_d \left( L {\bf Y}_{\bf 3}^{(4)} \right) + \gamma \tau_R H_d \left( L {\bf Y}_{\bf 3}^{(4)} \right).
\end{equation} 
In this case, the charged lepton mass matrix is 
\begin{align}
 M_E = \gamma Y^{(4)}_{3,3} \text{ diag}[\hat \alpha, \hat \beta, 1] 
\begin{pmatrix} \hat Y^{(4)}_{3,1} & 1 & \hat Y^{(4)}_{3,2} \\ \hat Y^{(4)}_{3,2} & \hat Y^{(4)}_{3,1} & 1 \\ 1 & \hat Y^{(4)}_{3,2} & \hat Y^{(4)}_{3,1} \end{pmatrix},
\label{eq:ME2}
\end{align}
where $\hat Y^{(4)}_{3,1(2)} \equiv Y^{(4)}_{3,1(2)}/Y^{(4)}_{3,3}$.
We separately analyze cases A and B since the charged lepton mass matrix affects the neutrino mass matrix through $V_L^e$ as we discussed above.

The neutrino mass matrix in this case is
\begin{equation}
m_\nu =  2 y_1 v_{T_1} (V^e_L)^T  \left[ \frac13
\begin{pmatrix} 2 Y^{(4)}_{3,1} & - Y^{(4)}_{3,3} & - Y^{(4)}_{3,2} \\ - Y^{(4)}_{3,3} & 2 Y^{(4)}_{3,2} & - Y^{(4)}_{3,1} \\ -Y^{(4)}_{3,2} & -Y^{(4)}_{3,1} & 2 Y^{(4)}_{3,3} \end{pmatrix} +
\hat{y_2} Y^{(4)}_{1} \begin{pmatrix} 1 & 0 & 0 \\ 0 & 0 & 1 \\ 0 & 1 & 0 \end{pmatrix} +
\hat{y_3} Y^{(4)}_{1'} \begin{pmatrix} 0 & 0 & 1 \\ 0 & 1 & 0 \\ 1 & 0 & 0 \end{pmatrix}
\right] V^e_L \ ,
\end{equation}
where $\hat y_{2,3} \equiv y_{2,3}/y_1$.

\subsection{Model (4)}

In this model, we chose $A_4$ singlets $1'',~1'$ and $1'$ for the triplets $T_1,~T_2$ and $H_u$ and 
the other assignments are the same as model (3).
Then the superpotential terms relevant to the neutrino masses are 
\begin{align}
w_\nu = & \ y_1 {\bf Y}_{\bf 3}^{(4)}(\tau) (L T_1 L)_{3} + y_2 Y_{1}^{(4)}(\tau) (L T_1 L)_{1} + y_3 Y_{1'}^{(4)}(\tau) (L T_1 L)_{1''} \nonumber \\
& + \lambda_2 H_u T_2 H_u + M_T T_1 T_2,
\end{align}
and it is the same as model (3) except for the $A_4$ structure.
The superpotential term relevant to the charged lepton masses is the same as model (3), 
and we also analyze cases A and B separately.

The neutrino mass matrix in this case is
\begin{equation}
m_\nu =  2 y_1 v_{T_1} (V^e_L)^T \left[ \frac13
\begin{pmatrix} 2 Y^{(4)}_{3,2} & - Y^{(4)}_{3,1} & - Y^{(4)}_{3,3} \\ - Y^{(4)}_{3,1} & 2 Y^{(4)}_{3,3} & - Y^{(4)}_{3,2} \\ -Y^{(4)}_{3,3} & -Y^{(4)}_{3,2} & 2 Y^{(4)}_{3,1} \end{pmatrix} +
\hat{y_2} Y^{(4)}_{1} \begin{pmatrix} 0 & 1 & 0 \\ 1 & 0 & 0 \\ 0 & 0 & 1 \end{pmatrix} +
\hat{y_3} Y^{(4)}_{1'} \begin{pmatrix} 1 & 0 & 0 \\ 0 & 0 & 1 \\ 0 & 1 & 0 \end{pmatrix}
\right] V^e_L \ ,
\end{equation}
where $\hat y_{2,3} \equiv y_{2,3}/y_1$.

\section{Numerical analysis}
In this section, we carry out the numerical analysis.
First, the free parameters in each model are scanned to search for regions in which the neutrino oscillation data 
can be accommodated.
Here we parametrize the Pontecorvo-Maki-Nakagawa-Sakata (PMNS) matrix $U_{PMNS}$, diagonalizing the neutrino mass matrix $m_\nu$, in terms of three mixing angles $\theta_{ij} (i,j=1,2,3; i < j)$, one CP violating Dirac phase $\delta_{CP}$,
and two Majorana phases $\{\alpha_{21}, \alpha_{32}\}$ as follows:
\begin{equation}
U_{PMNS} = 
\begin{pmatrix} c_{12} c_{13} & s_{12} c_{13} & s_{13} e^{-i \delta_{CP}} \\ 
-s_{12} c_{23} - c_{12} s_{23} s_{13} e^{i \delta_{CP}} & c_{12} c_{23} - s_{12} s_{23} s_{13} e^{i \delta_{CP}} & s_{23} c_{13} \\
s_{12} s_{23} - c_{12} c_{23} s_{13} e^{i \delta_{CP}} & -c_{12} s_{23} - s_{12} c_{23} s_{13} e^{i \delta_{CP}} & c_{23} c_{13} 
\end{pmatrix}
\begin{pmatrix} 1 & 0 & 0 \\ 0 & e^{i \frac{\alpha_{21}}{2}} & 0 \\ 0 & 0 & e^{i \frac{\alpha_{31}}{2}} \end{pmatrix},
\end{equation}
where $c_{ij}$ and $s_{ij}$ denote $\cos \theta_{ij}$ and $\sin \theta_{ij}$, respectively. 
Then we estimate the branching ratios of the doubly charged scalar boson focusing on the decays into 
the same sign charged lepton pairs using the allowed parameters explaining the neutrino oscillation data.

\subsection{Fitting neutrino oscillation data and relevant predictions}

Here we scan the free parameters in the models to fit the neutrino oscillation data.
In our analysis, we adopt experimentally allowed ranges for the mixing angles and mass squared differences at 3$\sigma$ range taken from ref.~\cite{Esteban:2018azc} as follows:
\begin{align}
& | \Delta m^2_{\rm atm}|=[2.431-2.622]\ ([2.413-2.606])\times 10^{-3}\ {\rm eV}^2 \quad \text{for NO(IO)},  \nonumber \\
& \Delta m^2_{\rm sol}=[6.79-8.01]\times 10^{-5}\ {\rm eV}^2 \quad \text{for both NO and IO},\nonumber \\
&\sin^2\theta_{13}=[0.02044-0.02437] \ ([0.02067-0.02461]) \quad \text{for NO(IO)},  \nonumber \\
& \sin^2\theta_{23}=[0.428-0.624] \ ([0.433-0.623]) \quad \text{for NO(IO)},\nonumber \\
& \sin^2\theta_{12}=[0.275-0.350] \quad \text{for both NO and IO},
\end{align}
where NO (IO) stands for normal (inverted) ordering for the neutrino masses.
Then the free parameters are scanned in the following range:
\begin{align}
& |\text{Re}[\tau]| \in [0, 0.5], \quad \text{Im}[\tau] \in [0.6, 2.0], \nonumber \\
& \hat y_{2,3} \in [-1.0, 1.0] \quad \text{for model (3) and (4)}.
\end{align}
The values of $2 y v_{T_1}/3$ and $2 y_1 v_{T_1}$ are fixed to provide the allowed range of $|\Delta m^2_{\rm atm}|$ 
using the value of $| \Delta m^2_{\rm atm}|$ as the input parameter.
\begin{center}
\begin{table}[tb]
\begin{tabular}{|c|c||c|c||c|c|c|} \hline
\multicolumn{2}{|c||}{~~~~Model~~~~} & ~~~(1)~~~ & ~~~(2)~~~ &~~~~~~~~~~~~ &~~~(3)~~~ & ~~~(4)~~~ \\ \hline
\multirow{4}{*}{~~~Normal~~~} &\multirow{2}{*}{~~~$1$ (red)~~~} & \multirow{2}{*}{$\times$} & \multirow{2}{*}{$\times$} 
          & Case A & Fig.~\ref{fig:M3C2NO} & Fig.~\ref{fig:M4C2NO} \\ \cline{5-7}
& & & & Case B &Fig.~\ref{fig:M3C2NOB} & Fig.~\ref{fig:M4C2NOB} \\ \cline{2-7}
&\multirow{2}{*}{$2$ (blue)} & \multirow{2}{*}{$\times$} & \multirow{2}{*}{$\times$} 
          & Case A & Fig.~\ref{fig:M3C2NO} &  Fig.~\ref{fig:M4C2NO} \\  \cline{5-7}
& & & & Case B & Fig.~\ref{fig:M3C2NOB} & Fig.~\ref{fig:M4C2NOB} \\  \hline
\multirow{4}{*}{~~~Inverted~~~} &\multirow{2}{*}{~~~$1$ (red)~~~} & \multirow{2}{*}{$\times$} & \multirow{2}{*}{$\times$} 
          & Case A & Fig.~\ref{fig:M3C2IO} & Fig.~\ref{fig:M4C2IO} \\ \cline{5-7}
& & & & Case B & Fig.~\ref{fig:M3C2IOB} & Fig.~\ref{fig:M4C2IOB} \\ \cline{2-7}
&\multirow{2}{*}{$2$ (blue)} & \multirow{2}{*}{$\times$} & \multirow{2}{*}{$\times$} 
          & Case A & Fig.~\ref{fig:M3C2IO} & Fig.~\ref{fig:M4C2IO} \\ \cline{5-7}
& & & & Case B & Fig.~\ref{fig:M3C2IOB} & Fig.~\ref{fig:M4C2IOB} \\  \hline
\end{tabular}
\caption{Summary of numerical analysis. Figure number is shown for the cases that the neutrino oscillation data 
can be fitted while cross mark ($\times$) stands for the case that the data can not be fitted. 
In Model column, $1$ (red) and $2$ (blue) 
corresponds to the models with Eq.~~\eqref{eq:alphabeta1} and \eqref{eq:alphabeta2}. Colors are used in figures 
to represent $1$ and $2$.}
\label{tab:summary}
\end{table}
\end{center}
The results of fitting the neutrino oscillation data are summarized in Table \ref{tab:summary}.
It should be noticed here that we assume $\hat y_{2}$ and $\hat y_{3}$ to be real values to simplify our analysis although 
these can be complex in general. 
This assumption can be justified by requiring that CP symmetry in the lepton sector is only violated through modular forms.
If we take complex $\hat y_{2}$ and $\hat y_{3}$, our predictions will be modified, in particular, by those for Dirac and 
Majorana phases.

For parameters accommodating the neutrino oscillation data, we compute the Jarlskog invariant, $J_{CP}$, 
which is 
given by the PMNS matrix elements $U_{\alpha i}$:
\begin{equation}
J_{CP} = \text{Im} [U_{e1} U_{\mu 2} U_{e 2}^* U_{\mu 1}^*] = s_{23} c_{23} s_{12} c_{12} s_{13} c^2_{13} \sin \delta_{CP}.
\end{equation}
The Majorana phases are also calculated via other invariants $I_1$ and $I_2$:
\begin{equation}
I_1 = \text{Im}[U^*_{e1} U_{e2}] = c_{12} s_{12} c_{13}^2 \sin \left( \frac{\alpha_{21}}{2} \right), \
I_2 = \text{Im}[U^*_{e1} U_{e3}] = c_{12} s_{13} c_{13} \sin \left( \frac{\alpha_{31}}{2}  - \delta_{CP} \right).
\end{equation}
We also calculate the effective mass for neutrinoless double beta decay given by
\begin{equation}
m_{ee} = |m_1 c^2_{12} c_{13}^2 + m_2 s^2_{12} c^2_{13} e^{i \alpha_{21}} + m_3 s^2_{13} e^{i(\alpha_{31} - 2 \delta_{CP})}|.
\end{equation}

\subsubsection{Model (1)}

In this model, the modulus $\tau$ is the only free parameter in the neutrino mass matrix except for overall factors associated with $y v_T$. 
For NO, it is found that we can fit the values of $| \Delta m^2_{\rm atm}|$, $| \Delta m^2_{\rm sol}|$ and $\sin^2 \theta_{12}$. 
However, the predicted values for the other mixing angles are $\sin^2 \theta_{23} \sim 0.8[0.5]$ and $\sin^2 \theta_{13} \sim 0.45[0.11]$ with Eq.~\eqref{eq:alphabeta1}[\eqref{eq:alphabeta2}] for charged lepton mass diagonalization, and they cannot be fully fitted to the observed data.
For IO, we find that only $| \Delta m^2_{\rm atm}|$ and $| \Delta m^2_{\rm sol}|$ can be consistent with the observed data.

\subsubsection{Model (2)}

This model is similar to model (1) except for the neutrino mass structure.
For NO, it is found that we can fit the values of $| \Delta m^2_{\rm atm}|$, $| \Delta m^2_{\rm sol}|$ and $\sin^2 \theta_{12}$. 
However, the predicted values for the other mixing angles are $\sin^2 \theta_{23} \sim 0.2$ and $\sin^2 \theta_{13} \sim 0.45$ 
for Eqs.~\eqref{eq:alphabeta1} and \eqref{eq:alphabeta2} solutions, 
and they cannot be fully fitted to the observed data.
For IO, we find that only $| \Delta m^2_{\rm atm}|$ and $| \Delta m^2_{\rm sol}|$ can be consistent with the observed data as 
in model (1).

\subsubsection{Model (3)}

\begin{figure}[tb]\begin{center}
\includegraphics[width=70mm]{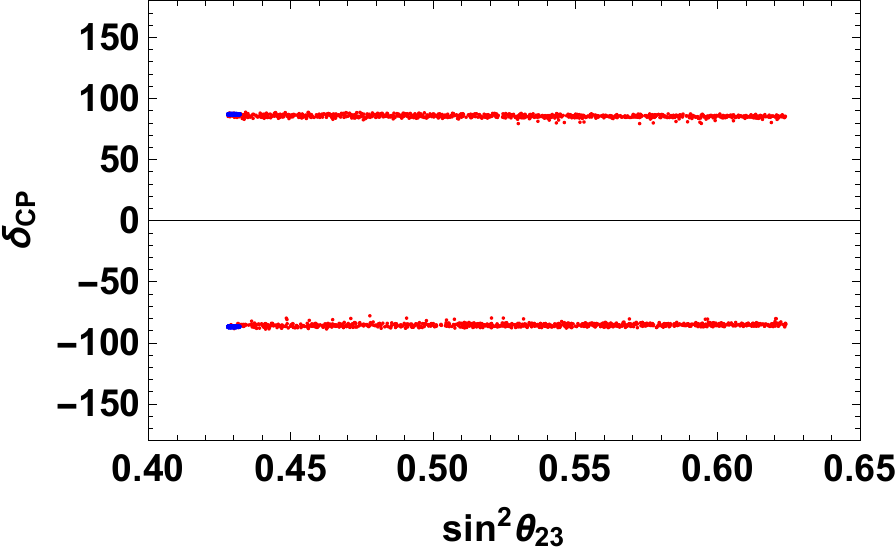} \
\includegraphics[width=70mm]{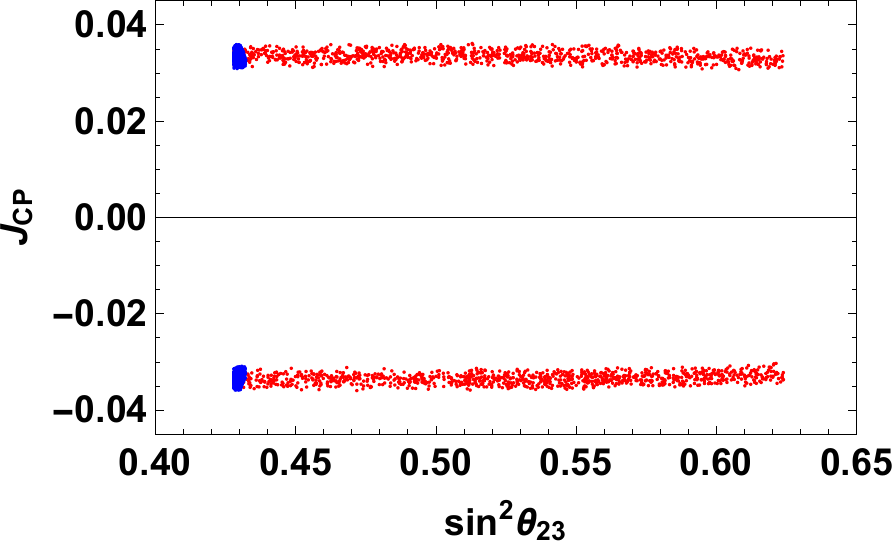} 
\includegraphics[width=70mm]{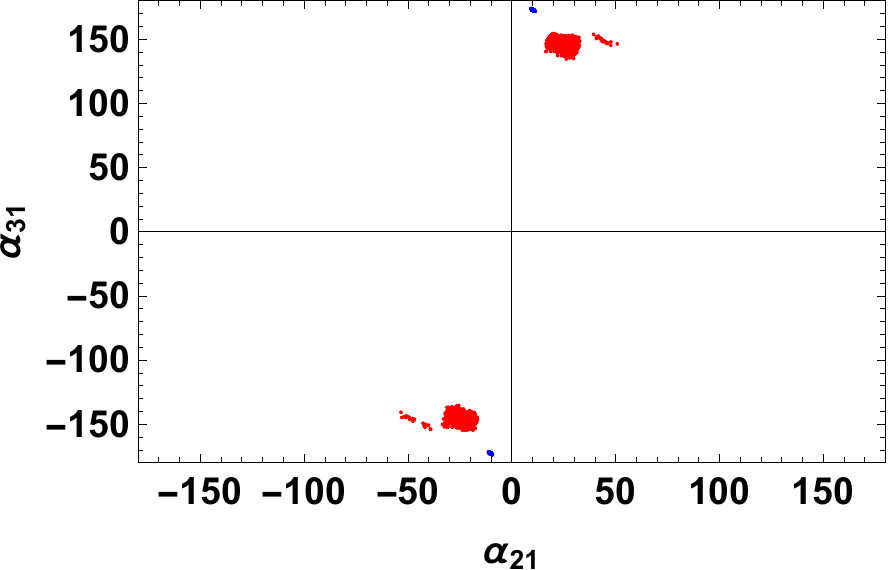} \
\includegraphics[width=70mm]{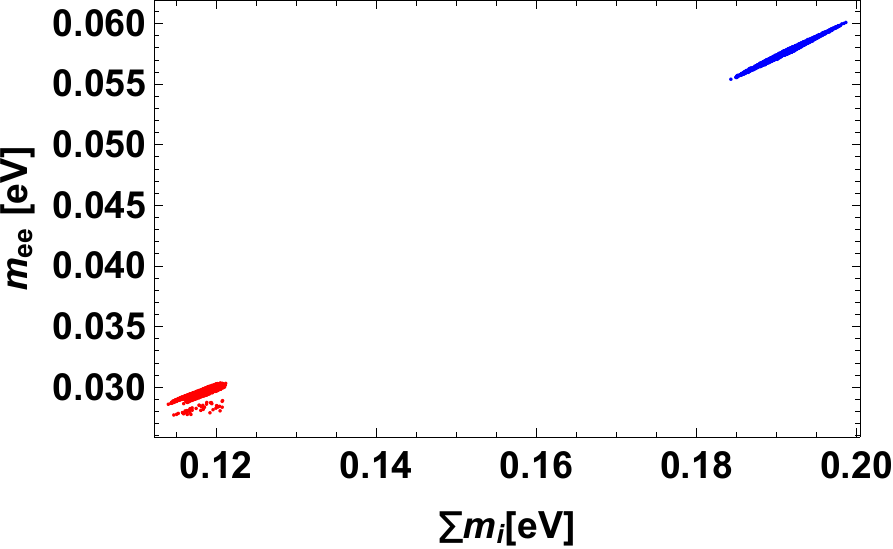}
\includegraphics[width=70mm]{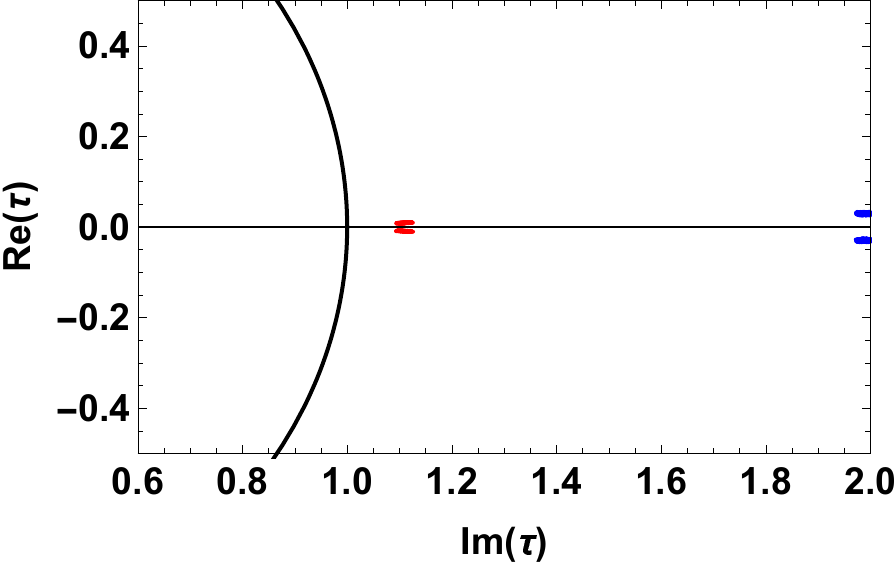} \
\includegraphics[width=70mm]{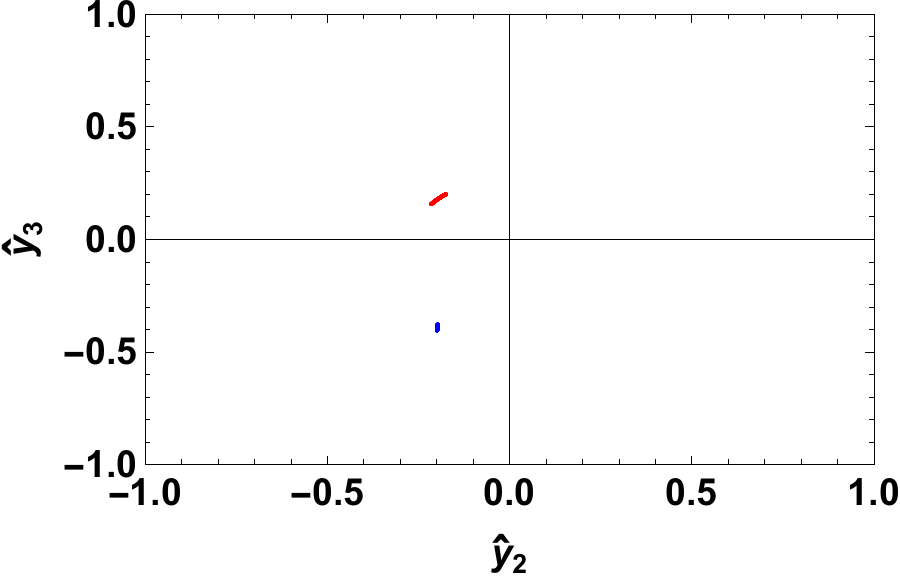}
\caption{Predictions in model (3) case A for NO. The top-left panel: predicted values on $\{\sin^2 \theta_{23}, \delta_{CP} \}$ plane. 
The top-right panel: predicted values on $\{\sin^2 \theta_{23}, J_{CP} \}$ plane. The center-left panel: predicted values on $\{\alpha_{21}, \alpha_{31} \}$ plane. 
The center-right panel: predicted values on $\{m_1, m_{ee}\}$ plane. 
The bottom-left panel: allowed region for real and imaginary part of $\tau$. Black solid curve indicates the fundamental domain of 
$|\tau|=1$. The bottom-right panel: allowed region for $\hat y_2$ and $\hat y_3$.
Here red[blue] points correspond to allowed parameter sets using Eq.~\eqref{eq:alphabeta1}[\eqref{eq:alphabeta2}] for $\hat \alpha$ and $\hat \beta$.}   
\label{fig:M3C2NO}\end{center}\end{figure}

\begin{figure}[tb]\begin{center}
\includegraphics[width=70mm]{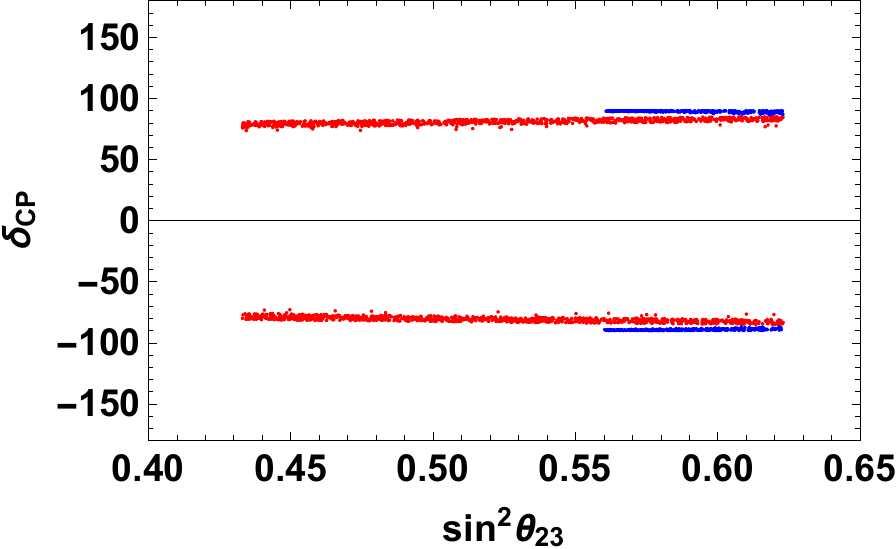} \
\includegraphics[width=70mm]{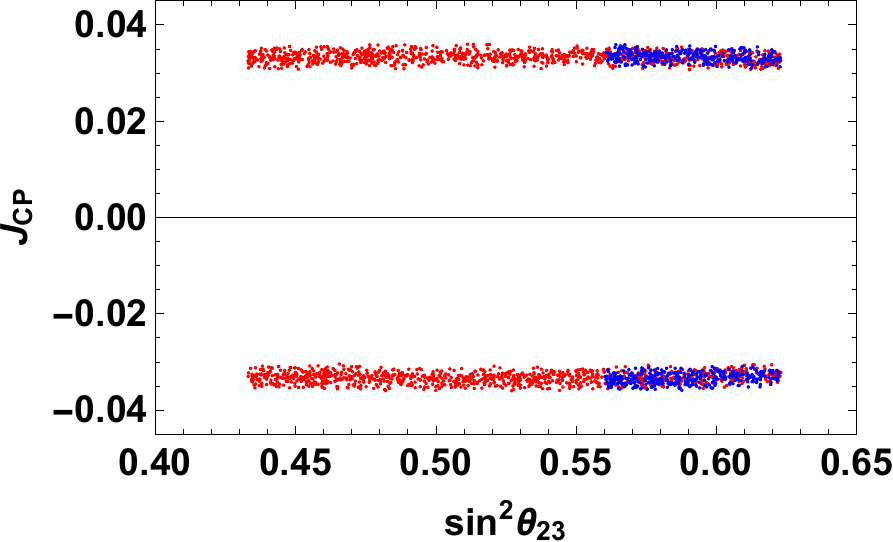} 
\includegraphics[width=70mm]{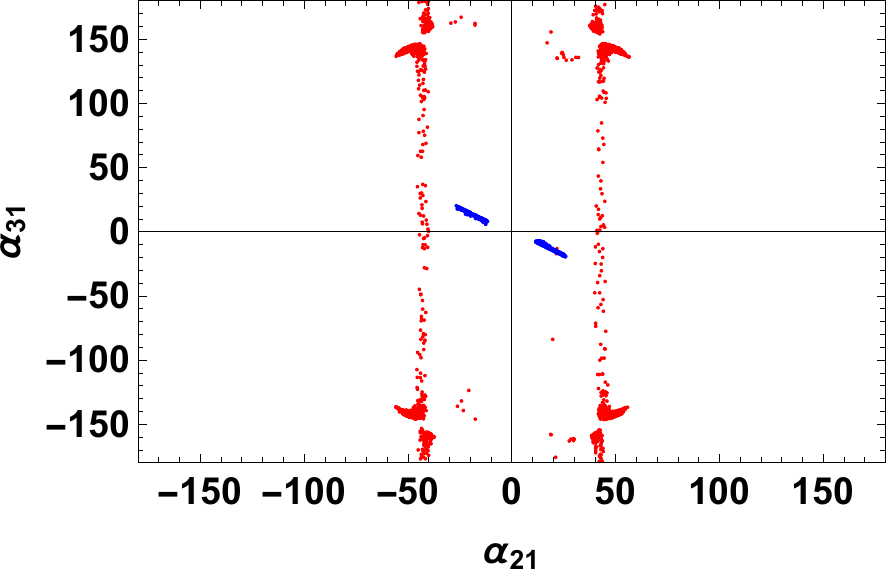} \
\includegraphics[width=70mm]{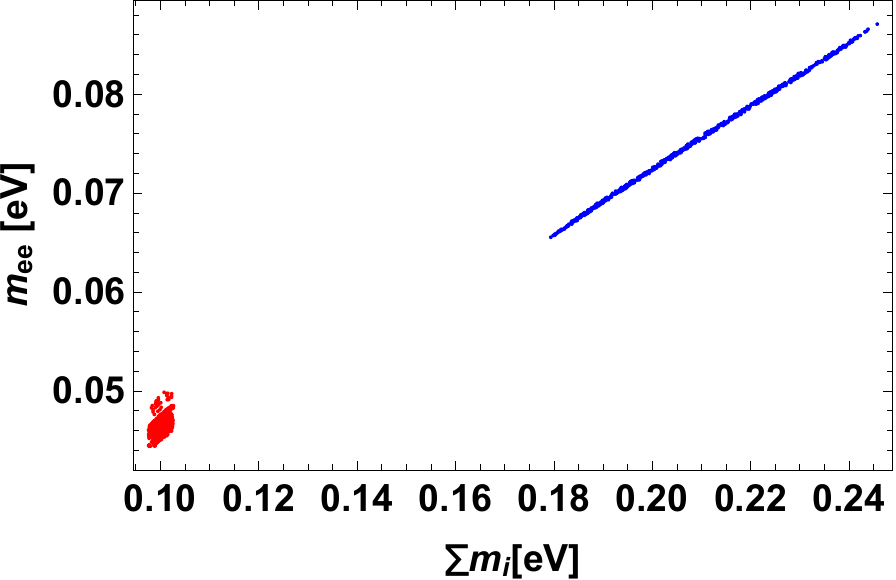}
\includegraphics[width=70mm]{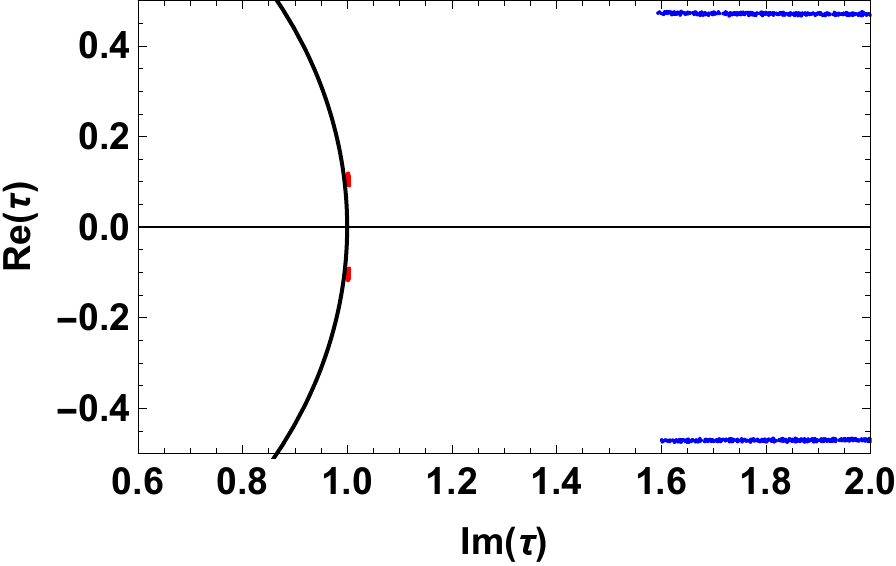} \
\includegraphics[width=70mm]{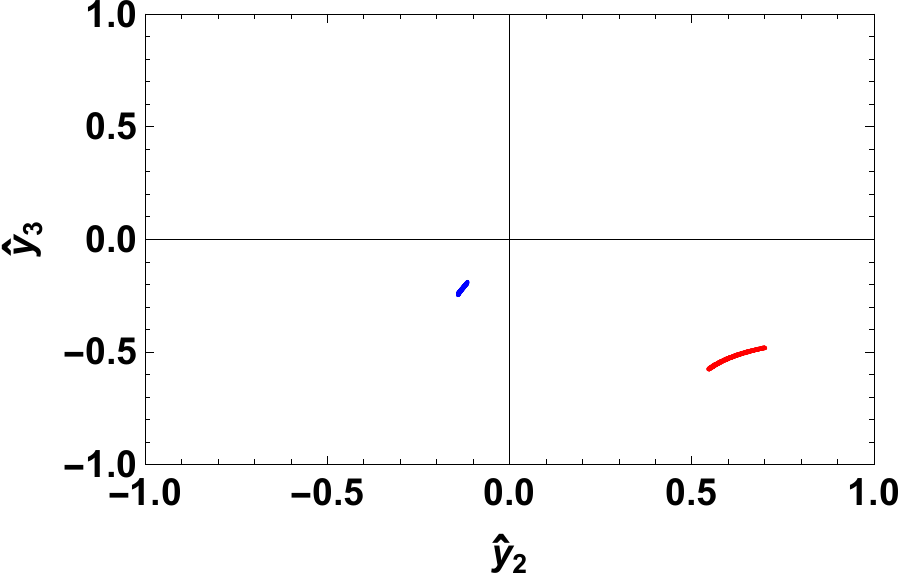}
\caption{The same plots as Fig.~\ref{fig:M3C2NO} in the case of model (3) case A for IO.}   
\label{fig:M3C2IO}
\end{center}\end{figure}

\begin{figure}[tb]\begin{center}
\includegraphics[width=70mm]{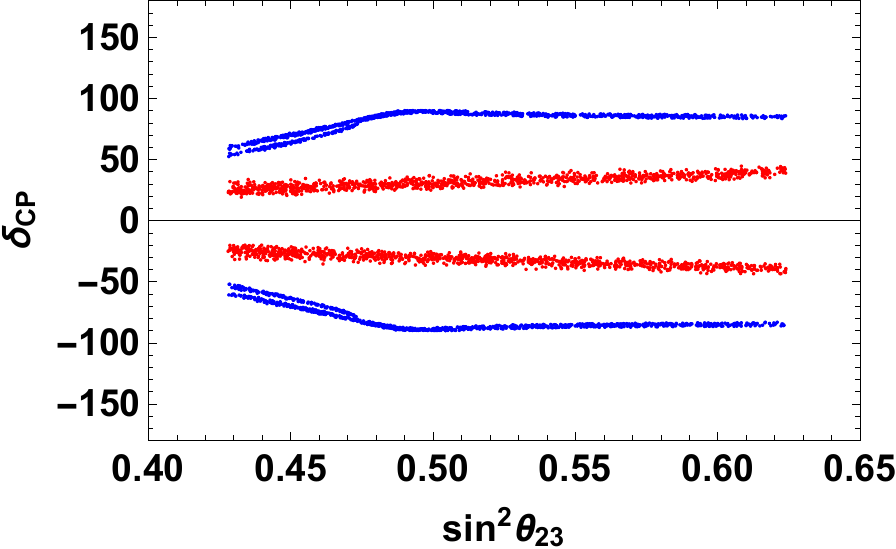} \
\includegraphics[width=70mm]{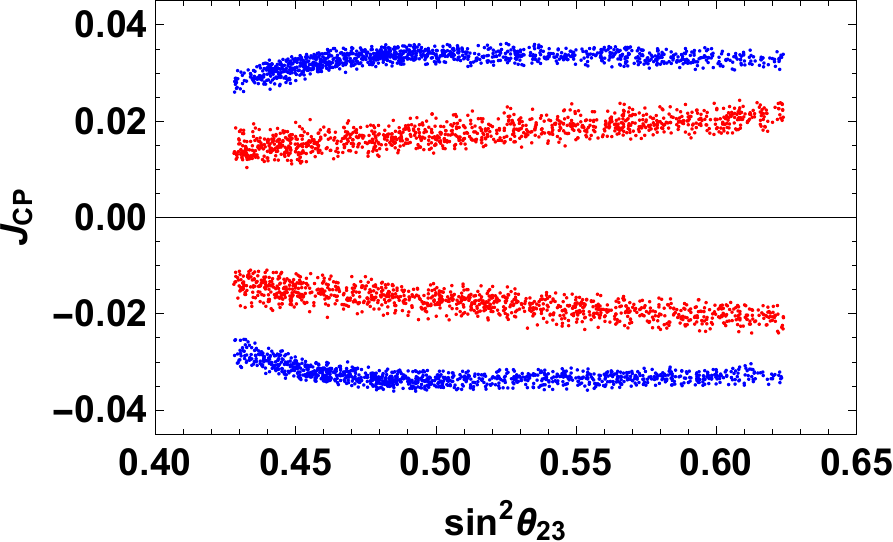} 
\includegraphics[width=70mm]{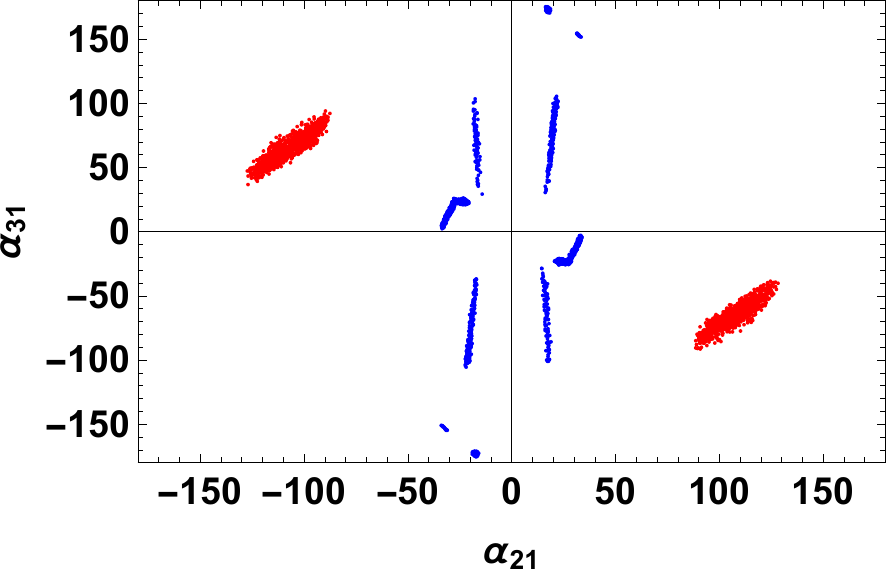} \
\includegraphics[width=70mm]{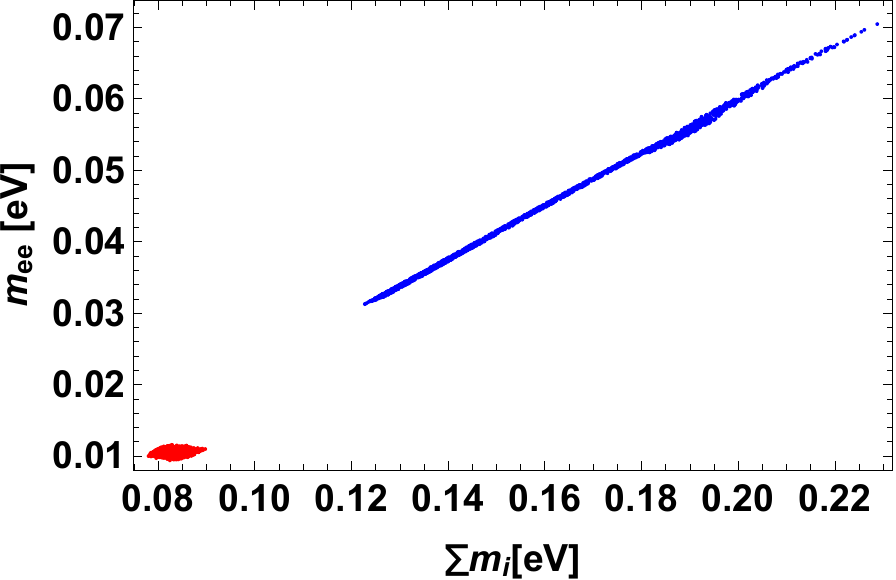}
\includegraphics[width=70mm]{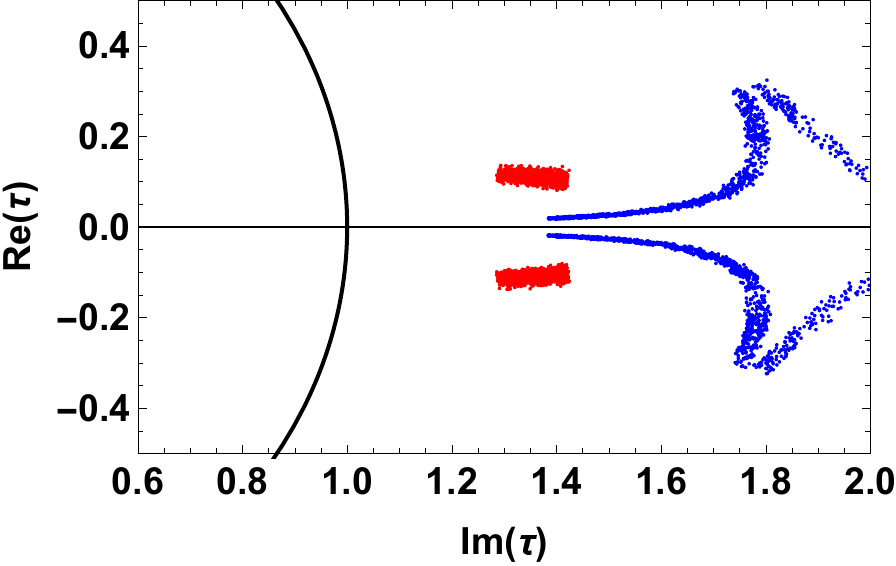} \
\includegraphics[width=70mm]{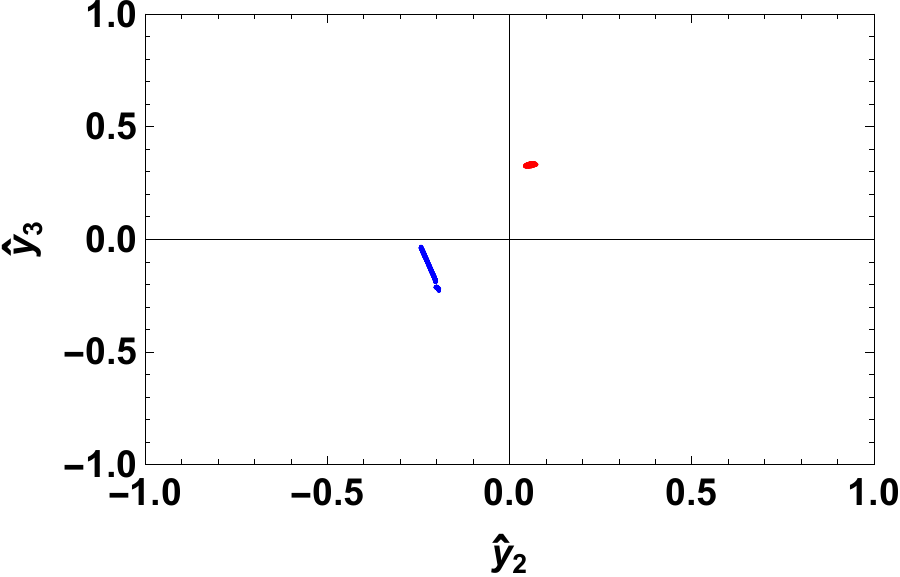}
\caption{The same plots as Fig.~\ref{fig:M3C2NO} in the case of Model (3) case B for NO. 
}   
\label{fig:M3C2NOB}\end{center}\end{figure}

\begin{figure}[tb]\begin{center}
\includegraphics[width=70mm]{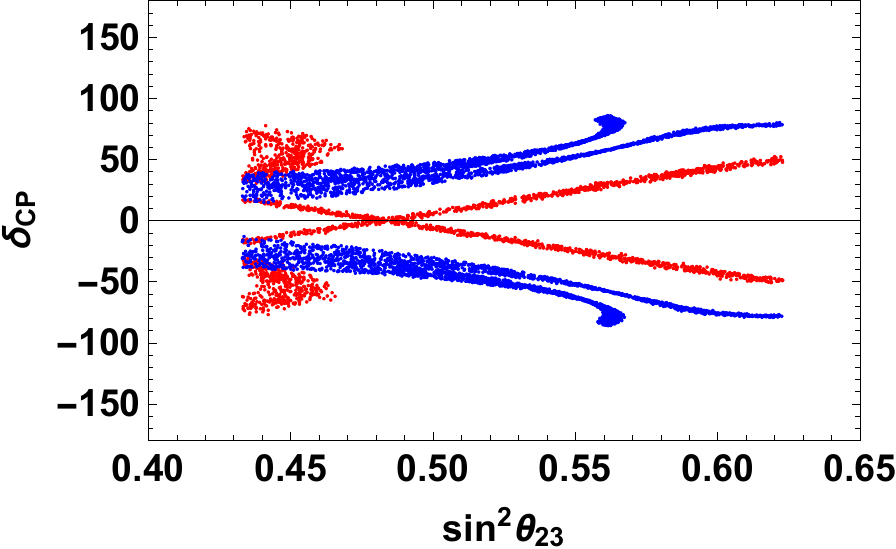} \
\includegraphics[width=70mm]{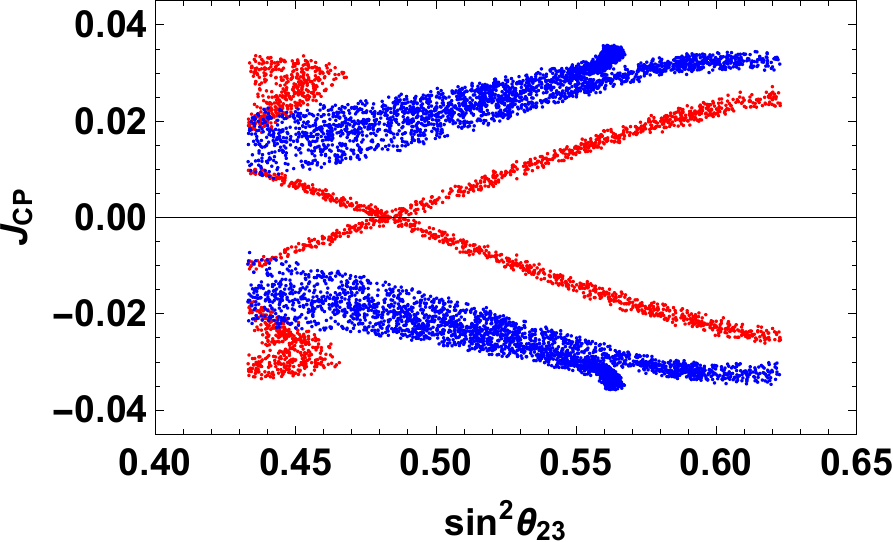} 
\includegraphics[width=70mm]{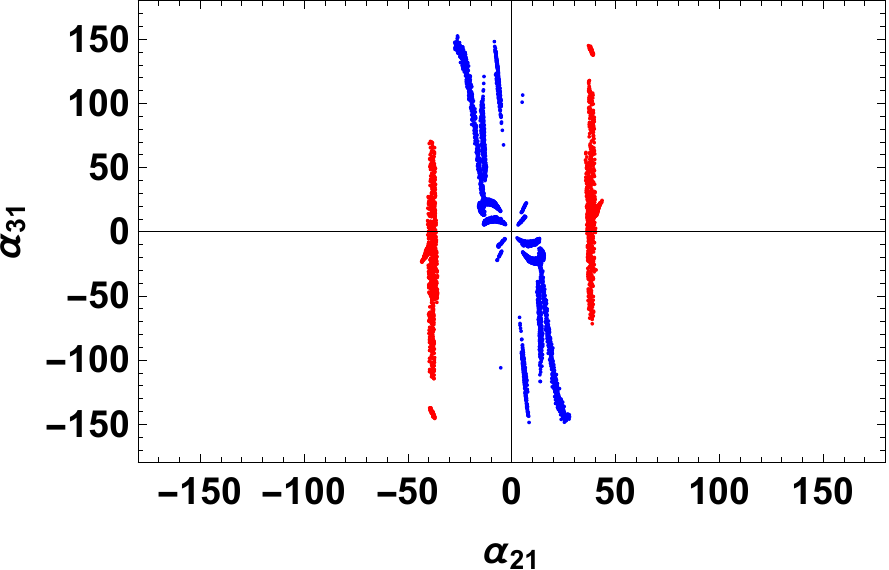} \
\includegraphics[width=70mm]{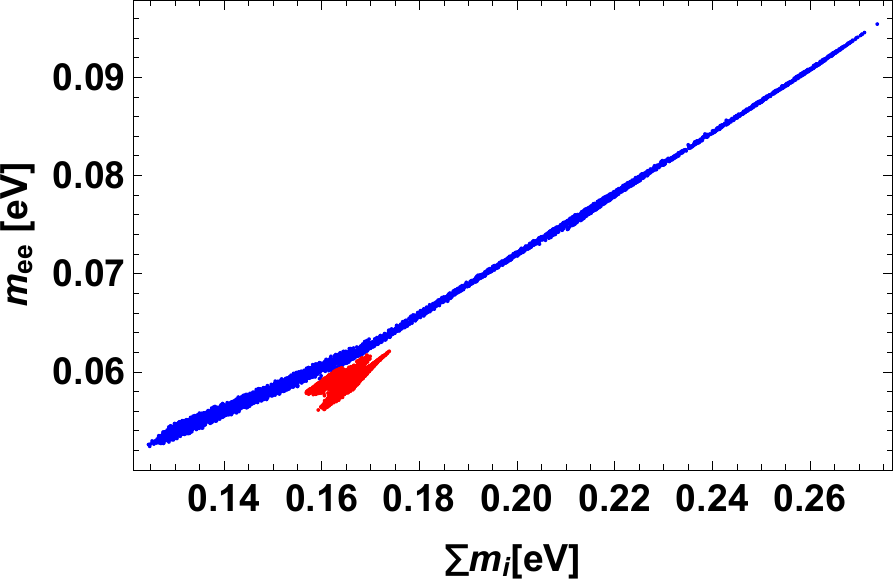}
\includegraphics[width=70mm]{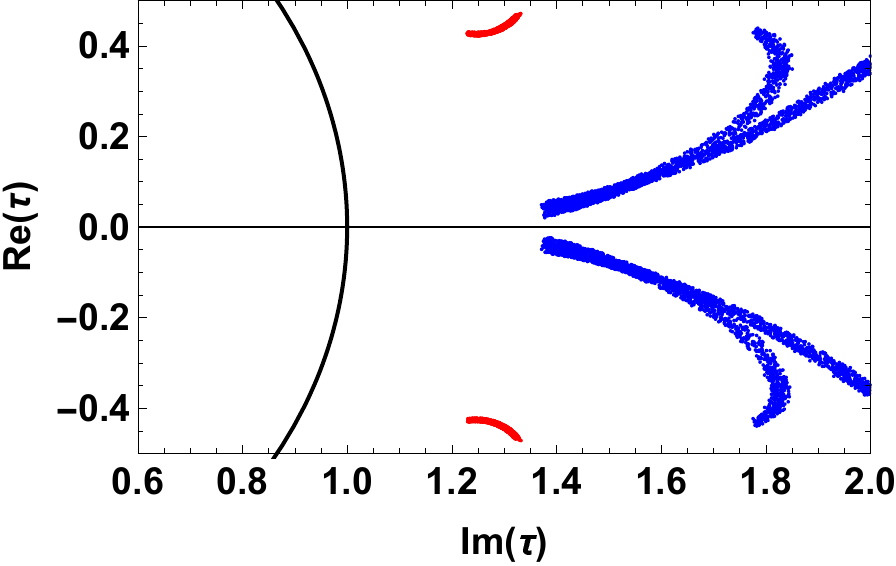} \
\includegraphics[width=70mm]{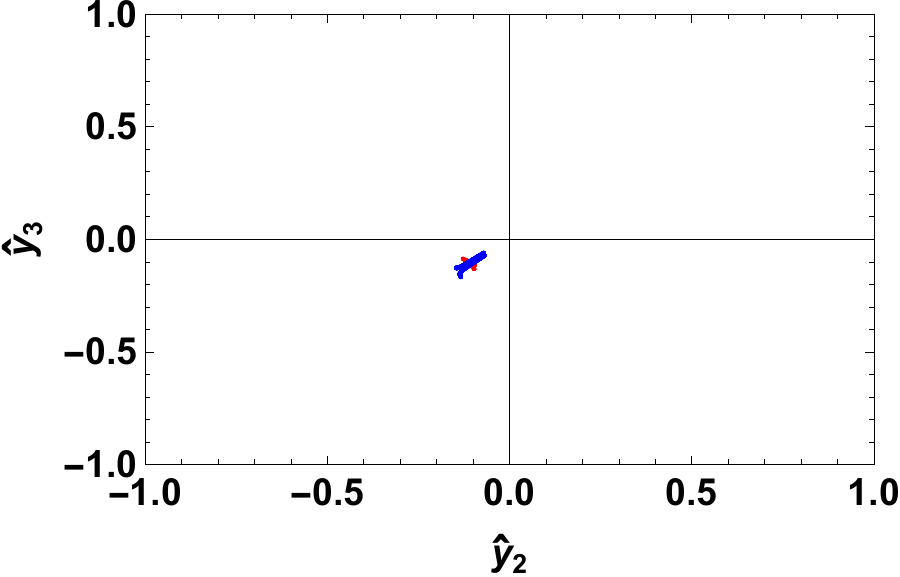}
\caption{The same plots as Fig.~\ref{fig:M3C2NO} in the case of model (3) case B for IO.}   
\label{fig:M3C2IOB}\end{center}\end{figure}

In model (3), we can fit the neutrino oscillation data due to the additional free parameters $\hat y_{2}$ 
and $\hat y_{3}$ which are absent in models (1) and (2). 
For cases A and B, the results are summarized as follows. 
In both cases, we find the parameter sets which can fit the neutrino oscillation data for both solutions of $\hat \alpha$ and $\hat \beta$ given by Eqs.~\eqref{eq:alphabeta1} and \eqref{eq:alphabeta2}.  \\ 
 {\bf Case A}: 
In Fig.~\ref{fig:M3C2NO}, we show our predictions (top and middle panels) and allowed parameter region 
(bottom panels) of our free parameters that satisfy 
neutrino oscillation data for NO.
Our predictions are shown on planes of $\{\sin^2 \theta_{23}, \delta_{CP} \}$, 
$\{\sin^2 \theta_{23}, J_{CP} \}$, $\{\alpha_{21}, \alpha_{31} \}$ and $\{\sum m_i, m_{ee}\}$ in the figure where red and blue points, respectively, correspond to the cases adopting Eqs.~\eqref{eq:alphabeta1} and \eqref{eq:alphabeta2} (we also use same color relation for the following plots). 
The predicted ranges of $\{\sin^2 \theta_{23}, \delta_{CP}\}$ are approximately $ \{[0.43, 0.62], [\pm 80^\circ, \pm 90^\circ] \}$ for Eq.~\eqref{eq:alphabeta1} and $\{0.43, [\pm 80^\circ,  \pm 90^\circ] \}$ for Eq.~\eqref{eq:alphabeta2}, respectively. 
We find that the predicted value of $\sin^2 \theta_{23}$ is restricted in the case of Eq.~\eqref{eq:alphabeta2}.
Then we also find several predicted regions on the $\{ \alpha_{21}, \alpha_{31}\}$ plane, which are located around $\{[0, \pm 60^\circ], [\pm140^\circ, \pm 180^\circ] \}$. 
The predicted ranges of $\{ \sum m_i, m_{ee}\}$ are approximately $\{ [0.11, 0.12],  [0.027, 0.031]\}$ eV for Eq.~\eqref{eq:alphabeta1} and $ \{ [0.183, 0.20],  [0.055, 0.060]\}$ eV for Eq.~\eqref{eq:alphabeta2}.
Furthermore the lightest neutrino mass is $m_1 \sim [0.03, 0.06]$ eV which has similar behavior as $m_{ee}$. 
Also, we find that the preferred regions of $\tau$ and $\hat y_{2,3}$ are different for solutions for $\hat \alpha$ and $\hat \beta$ as shown in bottom panels of the figure. 

For IO, the predictions and allowed parameter region are given in Fig.~\ref{fig:M3C2IO}.
In this case the predicted ranges of $\{\sin^2 \theta_{23}, \delta_{CP}\}$ are approximately $ \{[0.43, 0.62], [\pm 80^\circ, \pm 90^\circ] \}$ in the case of Eq.~\eqref{eq:alphabeta1} and $\{[0.56, 0.62], [\pm 90^\circ, \pm 100^\circ] \}$ in 
the case of Eq.~\eqref{eq:alphabeta2}, respectively. 
We find $\alpha_{21}$ is preferred to be $\sim 50^\circ$ while $\alpha_{32}$ can have wider range in the 
case of Eq.~\eqref{eq:alphabeta1}. On the other hand,  $\{ \alpha_{21}, \alpha_{31} \}$  are within $[-30^\circ, 30^\circ]$, approximately showing correlation in the case of Eq.~\eqref{eq:alphabeta2}. 
The predicted ranges of $\{ \sum m_i, m_{ee}\}$ are approximately $\{ 0.1,  [0.046-0.050]\}$ eV for Eq.~\eqref{eq:alphabeta1} and $ \{ [0.18, 0.24],  [0.065, 0.088]\}$ eV for Eq.~\eqref{eq:alphabeta2}, respectively.
Furthermore the lightest neutrino mass is also to be in two regions $m_1 \sim 0.045$ and  $m_1 \sim [0.07, 0.09]$ eV, which show similar behavior as $m_{ee}$.  \\

{\bf Case B}: For NO, the predictions and allowed parameter region are given in Fig.~\ref{fig:M3C2NOB}.
The predicted ranges of $\{\sin^2 \theta_{23}, \delta_{CP}\}$ values are approximately $ \{[0.43, 0.62], [\pm 20^\circ, \pm 40^\circ] \}$ for Eq.~\eqref{eq:alphabeta1} and $\{[0.43, 0.62], [\pm 50^\circ , \pm 90^\circ] \}$ for Eq.~\eqref{eq:alphabeta2}. 
We also find several allowed regions on the $\{ \alpha_{21}, \alpha_{32}\}$ plane indicating correlations between the angles for the region adopting Eq.~\eqref{eq:alphabeta1}. 
The predicted ranges of $\{ \sum m_i, m_{ee}\}$ are approximately $\{ [0.08, 0.09],  [0.01-0.012]\}$ eV for Eq.~\eqref{eq:alphabeta1} and $ \{ [0.12, 0.23],  [0.032, 0.070]\}$ eV for Eq.~\eqref{eq:alphabeta2}, respectively.
Furthermore the lightest neutrino mass is also to be in two regions, $m_1 \sim 0.01$ and $m_1 \sim [0.03, 0.07]$ eV, which shows similar behavior as $m_{ee}$. 

For IO, the predictions and allowed parameter region are given in Fig.~\ref{fig:M3C2IOB}.
The predicted ranges of $\{\sin^2 \theta_{23}, \delta_{CP}\}$ are approximately $ \{[0.43, 0.62], [- 80^\circ , 80^\circ] \}$ for Eq.~\eqref{eq:alphabeta1} and $\{[0.42, 0.62], [-90^\circ, 90^\circ] \}$ for Eq.~\eqref{eq:alphabeta2} respectively. 
We also find several allowed regions on the $\{ \alpha_{21}, \alpha_{31}\}$ plane within approximately $\pm 50^\circ$ where the region is more limited for Eq.~\eqref{eq:alphabeta1} around $\sim \pm 40^\circ$. 
The predicted ranges of $\{ \sum m_i, m_{ee}\}$ are approximately $\{ [0.16, 0.17], [0.056, 0.062]]\}$ eV for Eq.~\eqref{eq:alphabeta1} and $ \{ [0.13, 0.27], [0.052, 0.094]\}$ eV for Eq.~\eqref{eq:alphabeta2}, respectively.
Furthermore the lightest neutrino mass is also to be in two regions $m_1 \sim [0.055, 0.062]$ and $m_1 \sim [0.052, 0.094]$ eV,  which have similar behavior as $m_{ee}$.

\subsubsection{Model (4)}

\begin{figure}[tb]\begin{center}
\includegraphics[width=70mm]{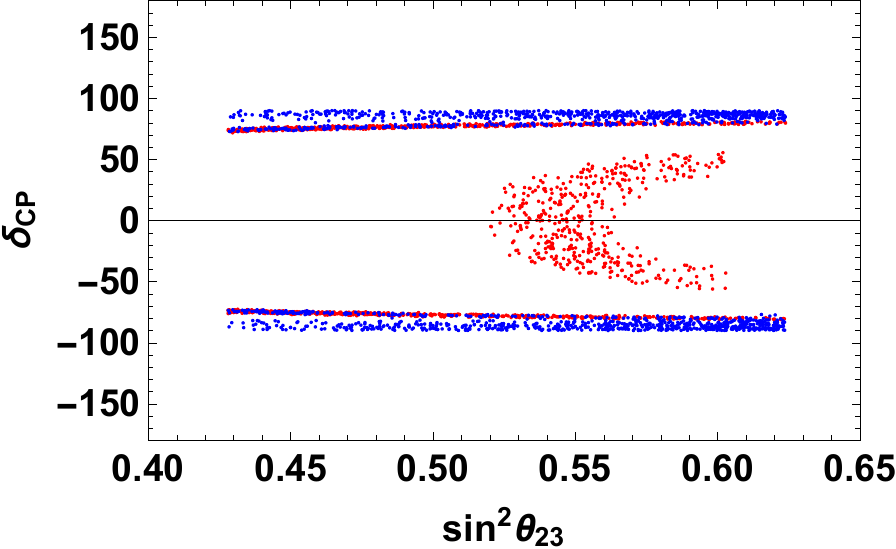} \
\includegraphics[width=70mm]{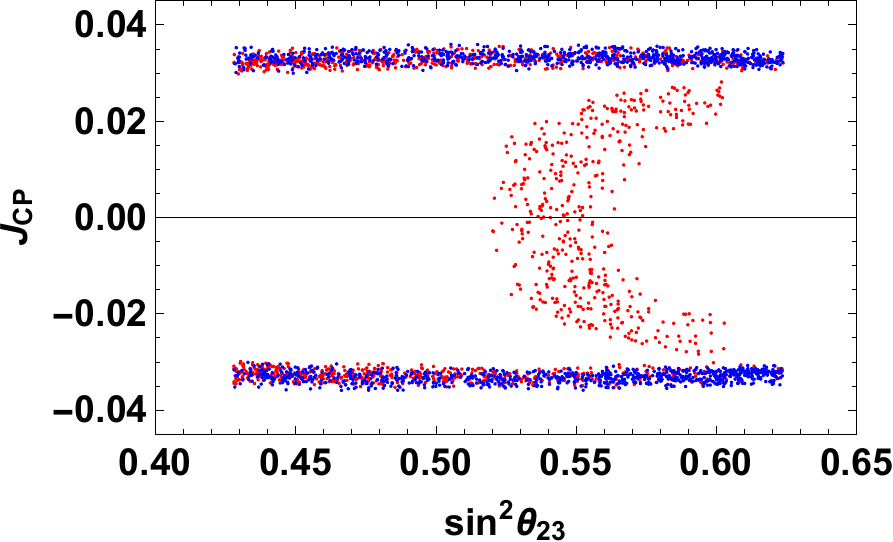} 
\includegraphics[width=70mm]{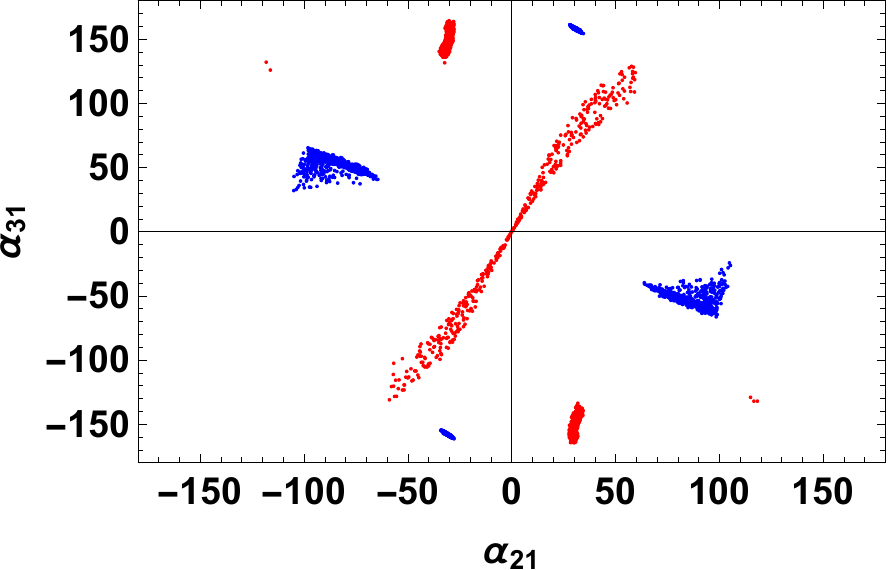} \
\includegraphics[width=70mm]{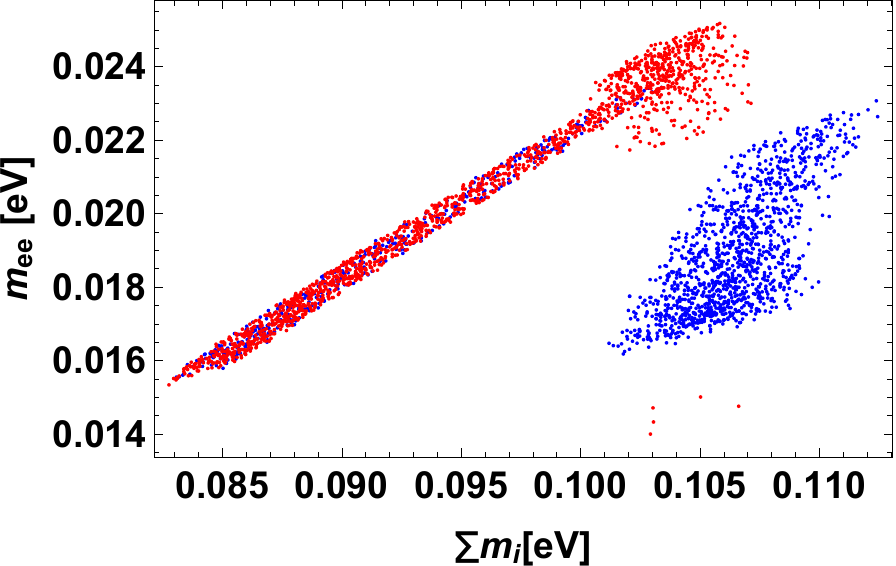}
\includegraphics[width=70mm]{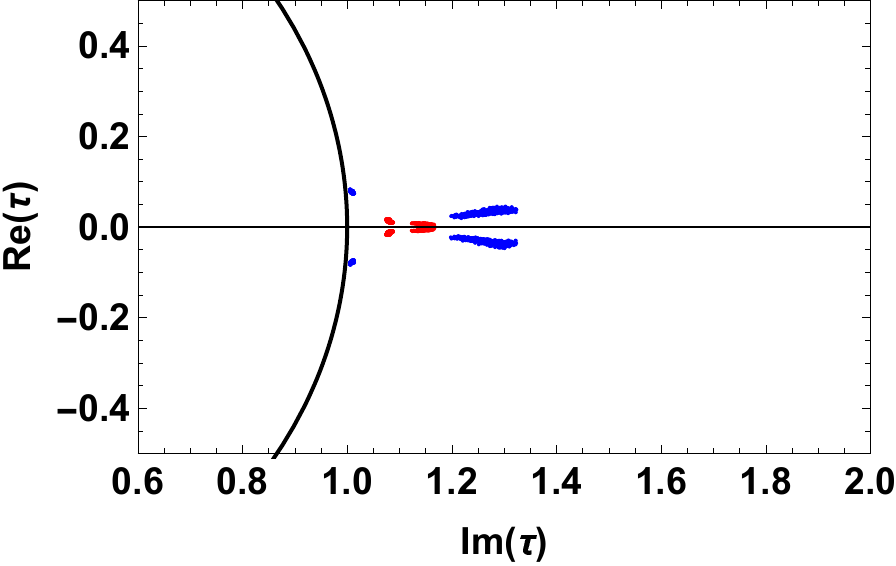} \
\includegraphics[width=70mm]{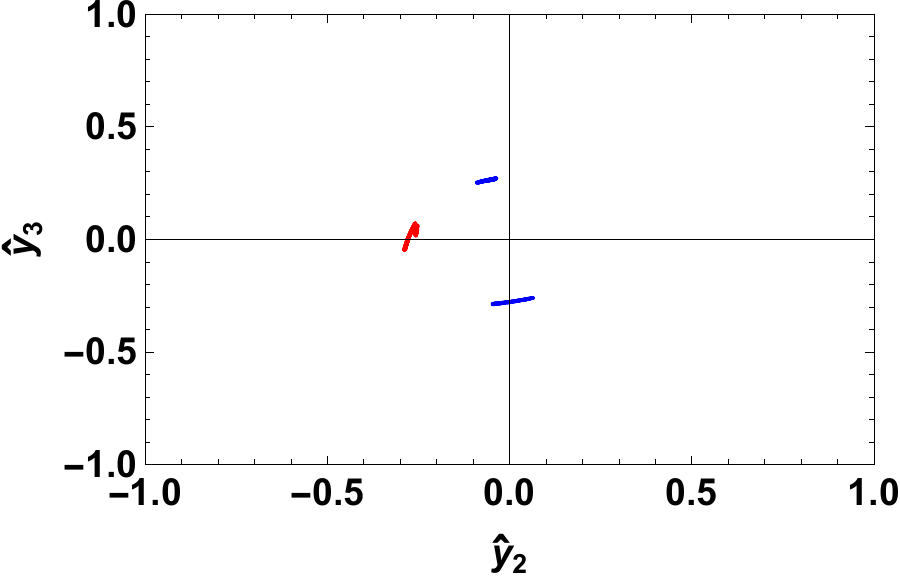}
\caption{The same plots as Fig.~\ref{fig:M3C2NO} in the case of model (4) case A for NO.}   
\label{fig:M4C2NO}\end{center}\end{figure}

\begin{figure}[tb]\begin{center}
\includegraphics[width=70mm]{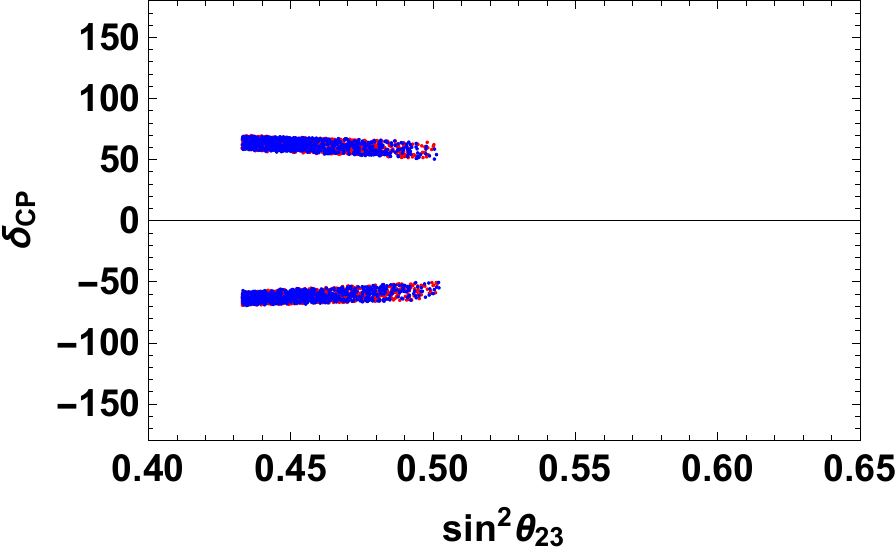} \
\includegraphics[width=70mm]{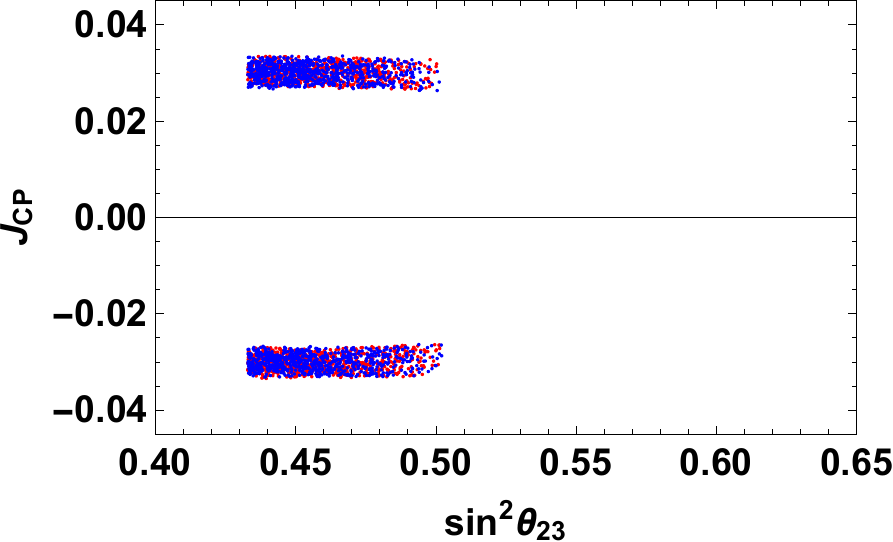} 
\includegraphics[width=70mm]{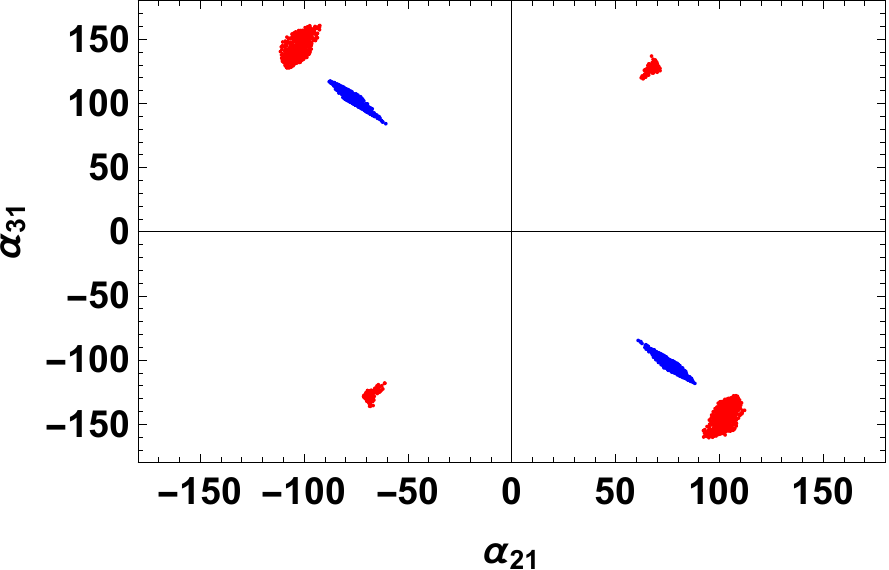} \
\includegraphics[width=70mm]{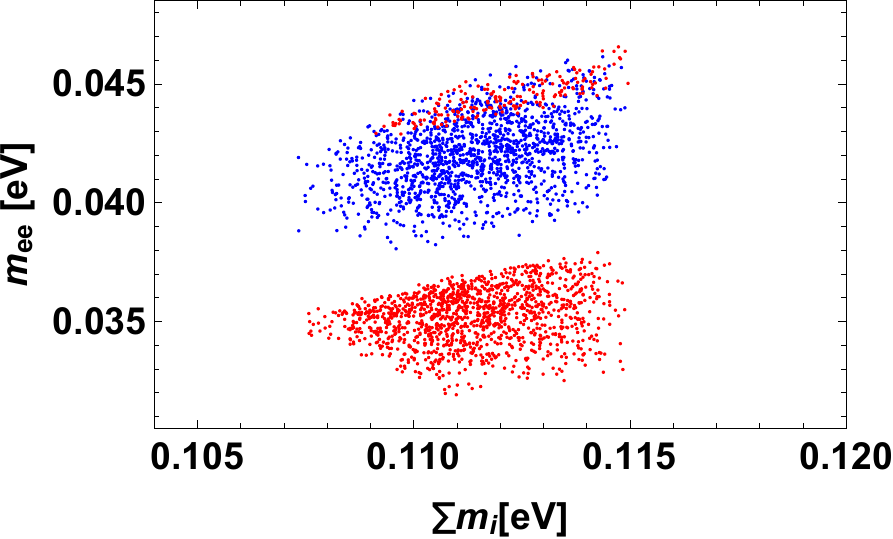}
\includegraphics[width=70mm]{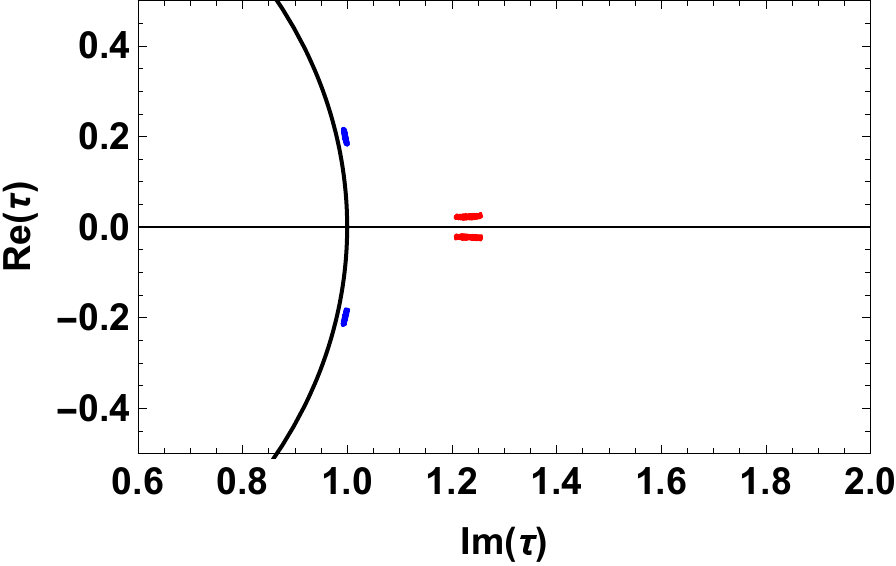} \
\includegraphics[width=70mm]{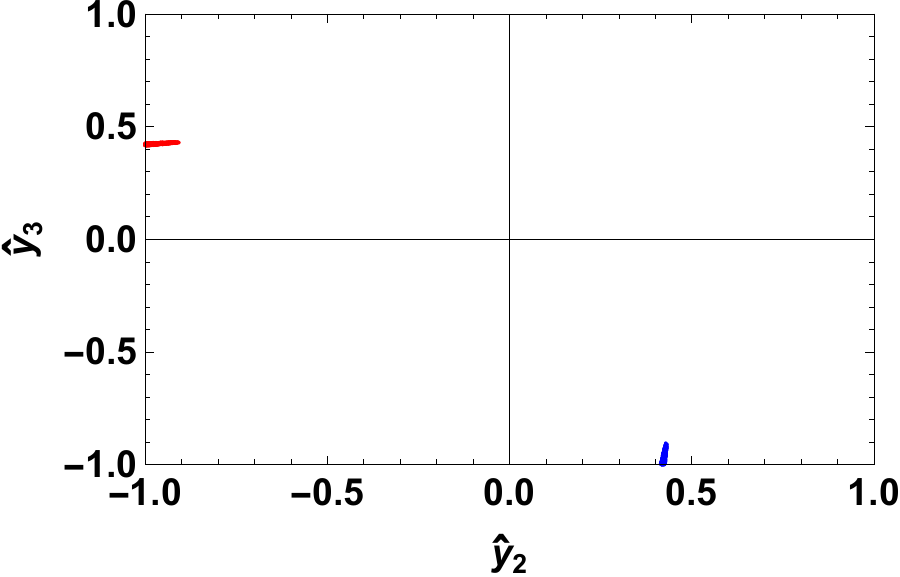}
\caption{The same plots as Fig.~\ref{fig:M3C2NO} in the case of model (4) case A for IO.}   
\label{fig:M4C2IO}\end{center}\end{figure}

\begin{figure}[tb]\begin{center}
\includegraphics[width=70mm]{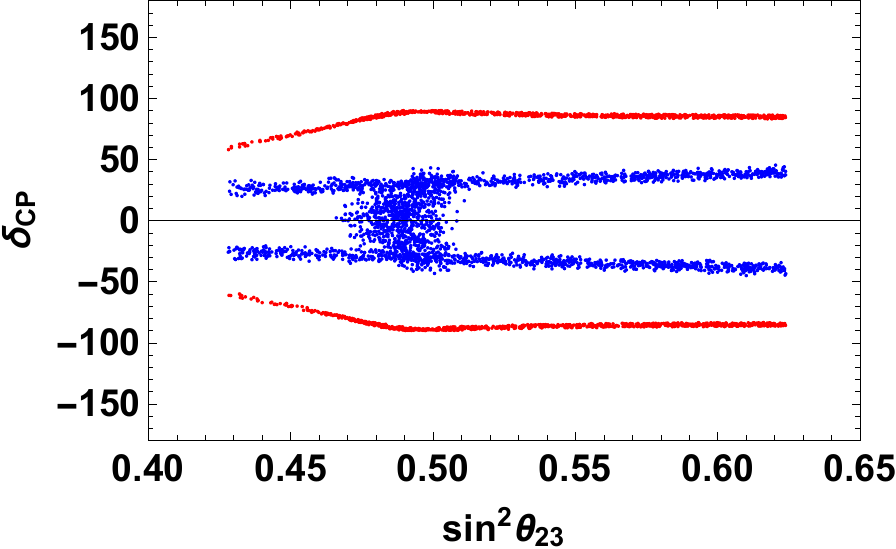} \
\includegraphics[width=70mm]{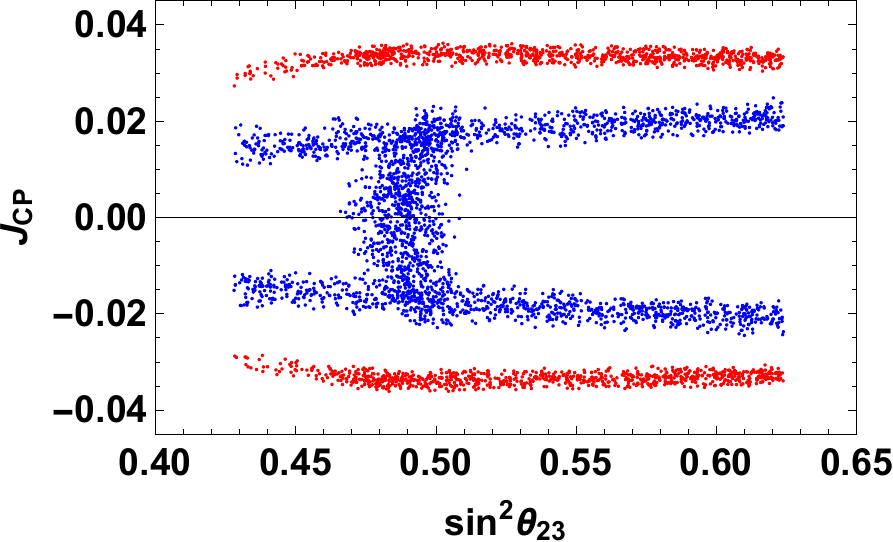} 
\includegraphics[width=70mm]{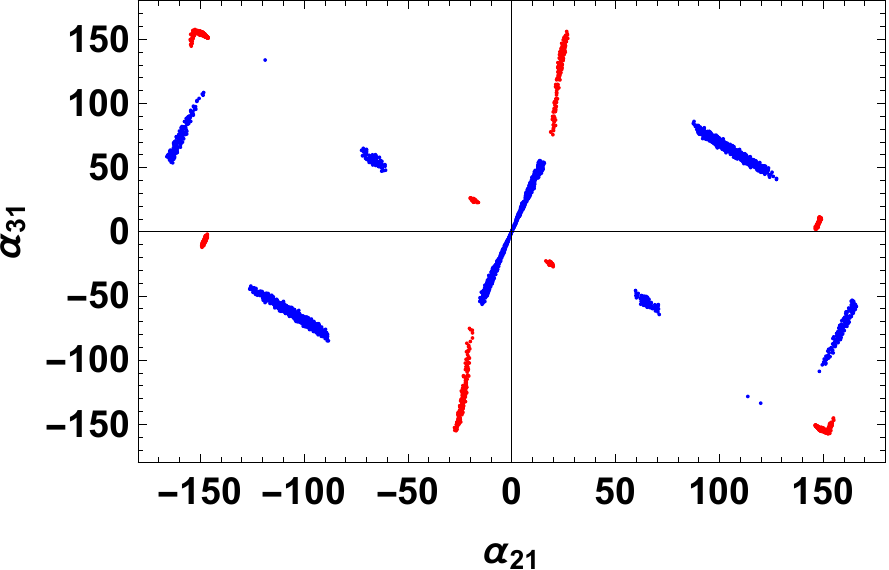} \
\includegraphics[width=70mm]{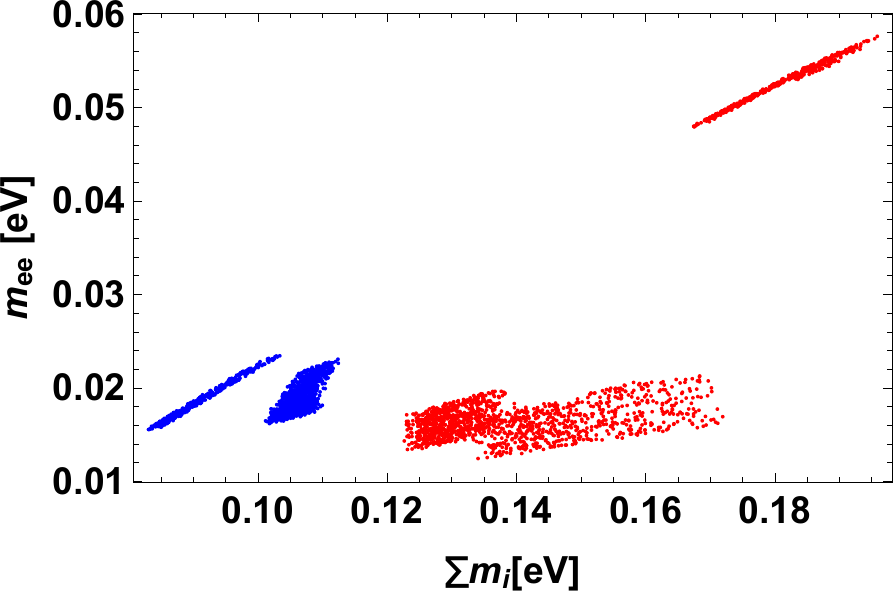}
\includegraphics[width=70mm]{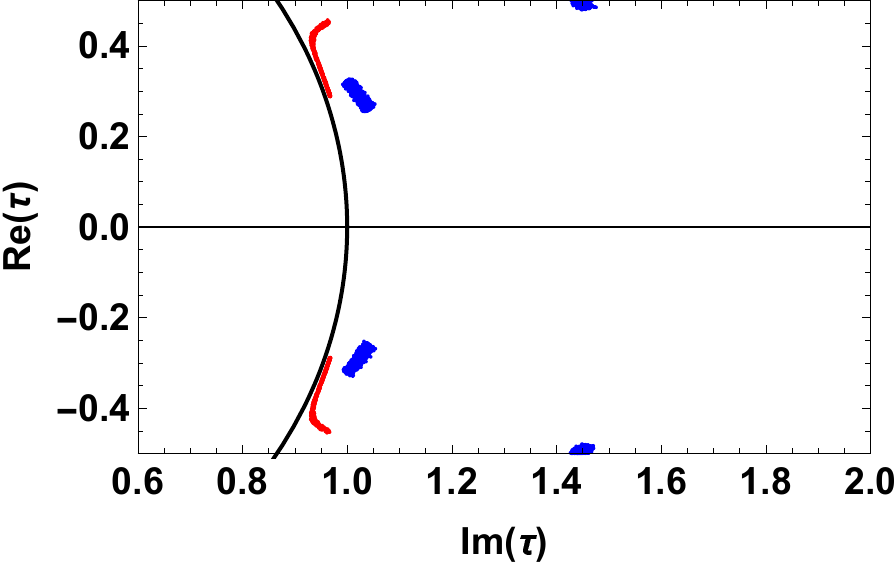} \
\includegraphics[width=70mm]{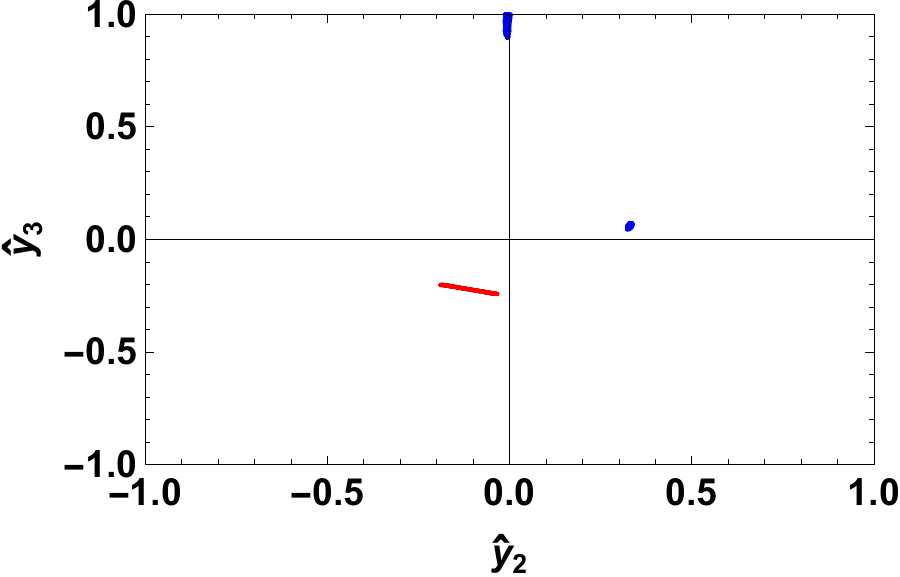}
\caption{The same plots as Fig.~\ref{fig:M3C2NO} in the case of model (4) case B for NO.}   
\label{fig:M4C2NOB}\end{center}\end{figure}

\begin{figure}[tb]\begin{center}
\includegraphics[width=70mm]{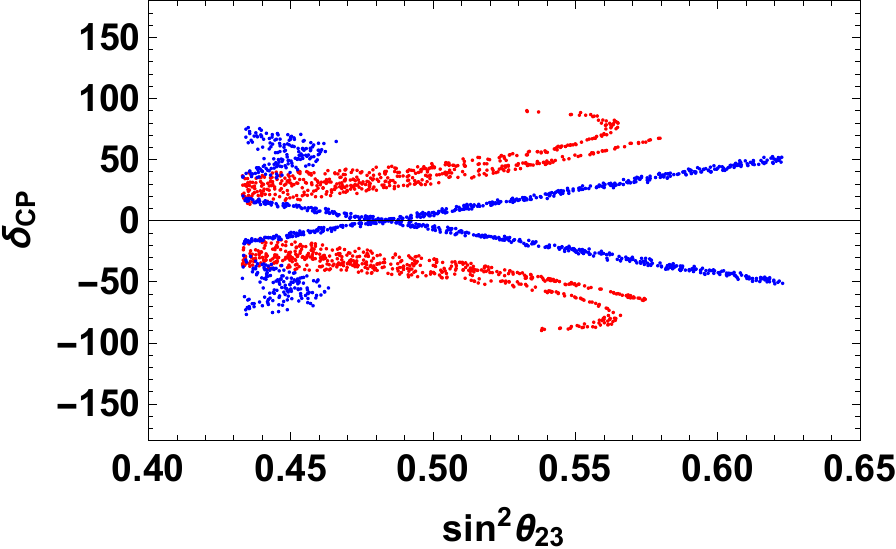} \
\includegraphics[width=70mm]{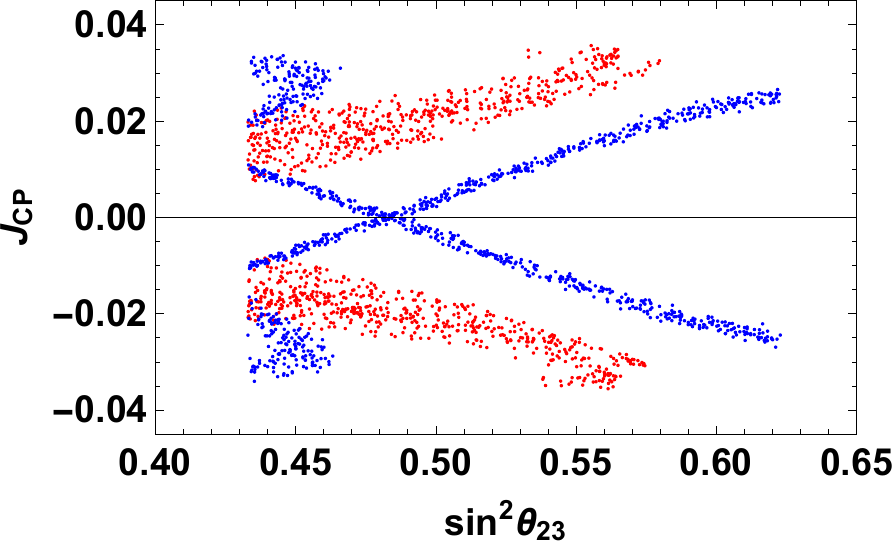} 
\includegraphics[width=70mm]{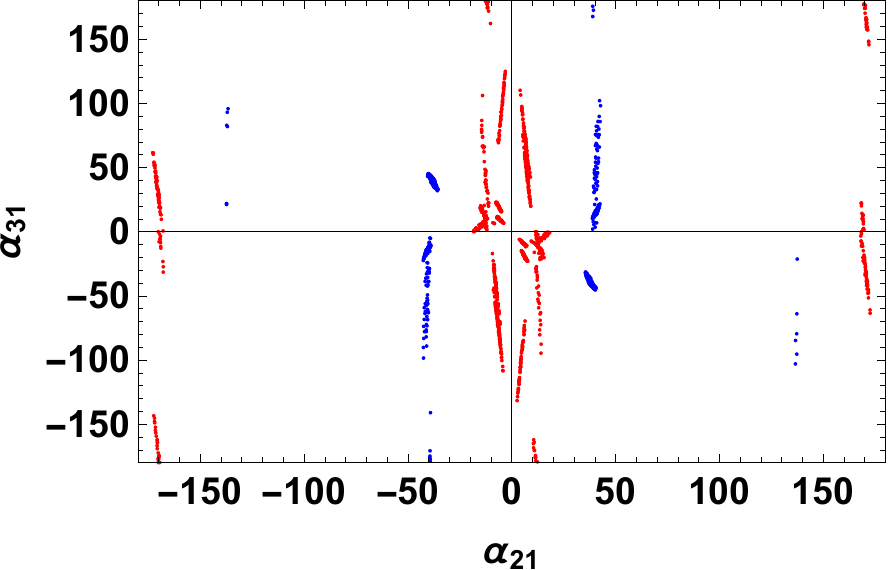} \
\includegraphics[width=70mm]{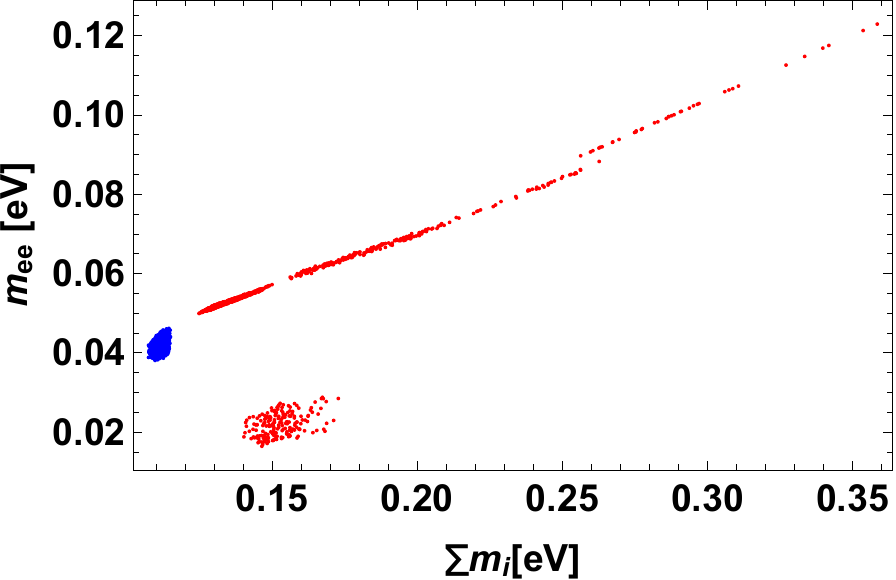}
\includegraphics[width=70mm]{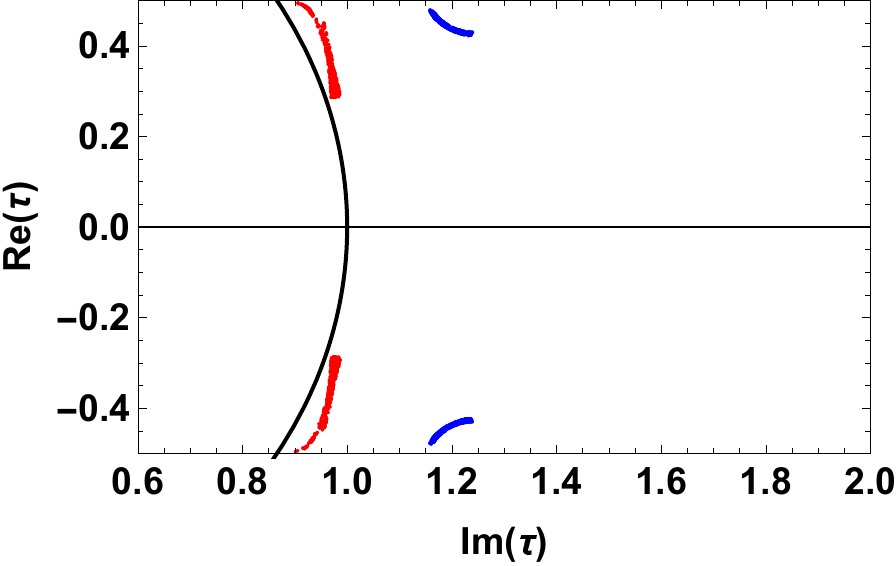} \
\includegraphics[width=70mm]{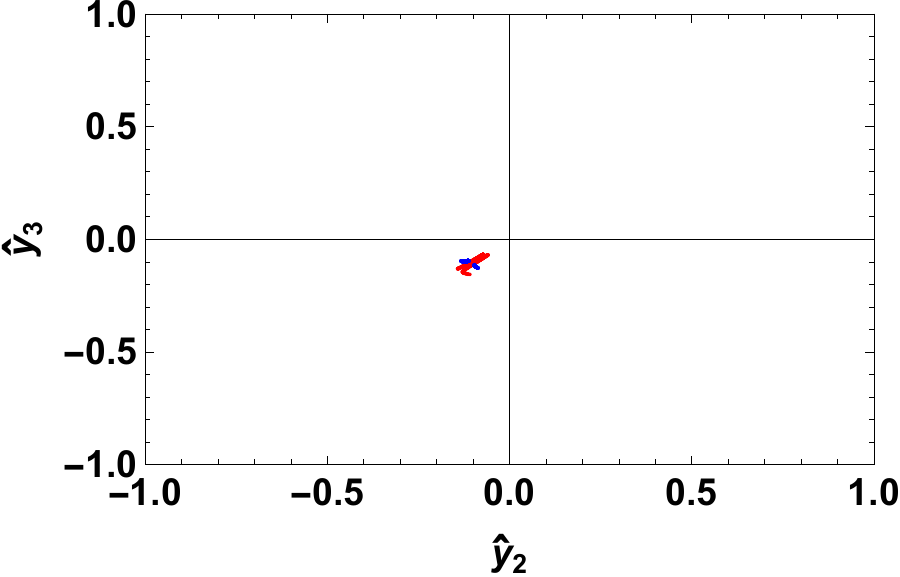}
\caption{The same plots as Fig.~\ref{fig:M3C2NO} in the case of model (4) case B for IO.}   
\label{fig:M4C2IOB}\end{center}\end{figure}

Model (4) also can accommodate the neutrino oscillation data due to the additional parameters for both solutions of $\hat \alpha$ and $\hat \beta$ given by Eqs.~\eqref{eq:alphabeta1} and \eqref{eq:alphabeta2}. The results for cases A and B are as follows. \\
{\bf Case A}: For NO,  the predictions and allowed parameter region are given in Fig.~\ref{fig:M4C2NO}.
In this case we obtain several regions for $\{\sin^2 \theta_{23}, \delta_{CP}\}$ for both Eq.~\eqref{eq:alphabeta1} and Eq.~\eqref{eq:alphabeta2} cases as shown in the top panels of the figure where we omit the detailed explanation. 
We also find several allowed regions on the $\{ \alpha_{21}, \alpha_{31}\}$ plane. 
In this case $\{\sum m_i, m_{ee} \}$ values are also found in the region within an approximate range of $\sim \{[0.082, 0.105], [0.015, 0.025] \}$ for Eq.~\eqref{eq:alphabeta1} and $\sim \{[0.101, 0.113], [0.016, 0.022] \}$ for Eq.~\eqref{eq:alphabeta2}.
Furthermore the lightest neutrino mass is also found to be in some regions within $m_1 \sim [0.015, 0.025]$ eV as similar to $m_{ee}$. 

For IO, the predictions and allowed parameter region are given in Fig.~\ref{fig:M4C2IO}. 
The predicted range of $\{\sin^2 \theta_{23}, \delta_{CP}\}$ is approximately $ \{[0.43, 0.50], [\pm 50^\circ , \pm 70^\circ] \}$ where the region is almost same for Eq.~\eqref{eq:alphabeta1} and Eq.~\eqref{eq:alphabeta2}. 
We also find several small predicted regions on the $\{ \alpha_{21}, \alpha_{31}\}$ plane within around $\alpha_{21} \sim [\pm 50^\circ, \mp 110^\circ] $ and $\alpha_{31} \sim [\pm 90^\circ, \mp 180^\circ] $. 
In this case $\{\sum m_i, m_{ee} \}$ values are also found in regions approximately within $\{ [0.107, 0.115], [0.032, 0.047] \}$.
Furthermore the lightest neutrino mass is also found to be within $m_1 \sim [0.032, 0.047]$ eV.\\
{\bf Case B}: For NO,  the predictions and allowed parameter region are given in Fig.~\ref{fig:M4C2NOB}.
In this case the predicted ranges of $\{\sin^2 \theta_{23}, \delta_{CP}\}$ are approximately $ \{[0.43, 0.62], [\pm 50^\circ, \pm 90^\circ] \}$ for Eq.~\eqref{eq:alphabeta1} and $\{[0.43, 0.62], [-40^\circ,  40^\circ] \}$ for Eq.~\eqref{eq:alphabeta2} respectively.  
We also find several allowed regions on the $\{ \alpha_{21}, \alpha_{31}\}$ plane as shown in the figure omitting detailed explanation. 
The predicted regions of $\{ \sum m_i, m_{ee}\}$ are approximately $\{ [0.085, 0.17], [0.01 0.025]]\}$ eV and $ \{ [0.17, 0.195], [0.05, 0.06]\}$ eV where the region with larger values is obtained from Eq.~\eqref{eq:alphabeta1}.
Furthermore, the lightest neutrino mass is also found to be in some regions within $m_1 \sim [0.01, 0.06]$ eV as similar to $m_{ee}$. 

For IO,  the predictions and allowed parameter region are given in Fig.~\ref{fig:M4C2IOB}.
The predicted ranges of $\{\sin^2 \theta_{23}, \delta_{CP}\}$ are approximately $ \{[0.43, 0.62], [\pm 10^\circ , \pm 90^\circ] \}$ for Eq.~\eqref{eq:alphabeta1} and $\{[0.43, 0.62], [-80^\circ, 80^\circ] \}$ for Eq.~\eqref{eq:alphabeta2} respectively. 
We also find several predicted regions on the $\{ \alpha_{21}, \alpha_{31}\}$ plane where $\alpha_{21}$ is restricted around $\sim [-50^\circ, 50^\circ] $ and $ [ \pm 140^\circ , \pm 180^\circ \}$ 
while $\alpha_{31}$ can be any value. 
In this case $\{\sum m_i, m_{ee} \}$ values are also found in some different regions approximately within $\{ [0.11, 0.36], [0.02, 0.12] \}$ where the region is more restricted for Eq.~\eqref{eq:alphabeta2} as 
$\sum m_i \sim 0.11$ and $m_{ee} \sim 0.04$.
Furthermore the lightest neutrino mass is also found to be in some regions within $m_1 \sim [0.02, 0.12]$ eV as similar to $m_{ee}$.

\subsection{Branching ratio of doubly charged scalar boson}

Here we calculate the BRs of the doubly charged scalar boson $\delta^{\pm \pm}$.
In the type-II seesaw model, $\delta^{\pm \pm} \to \ell^\pm \ell^\pm$ decay modes are induced via Yukawa couplings 
\begin{equation}
L \supset \frac{1}{2 v_{T_1}} \bar \ell^{c}_{L_i} (m_\nu)_{ij} \ell_{L_j} \delta^{++} + h.c.,
\end{equation}
where $m_\nu$ is the neutrino mass matrix. 
The doubly charged scalar also decays into the same sign $W$ boson pair through the gauge interaction which is proportional to $v_{T_1}$.
Then leptonic modes are dominant when $v_{T_1} < 10^{-4}$ GeV~\footnote{We can also have decay modes with other scalar bosons in triplet. They can be ignored when masses for components of triplet are degenerated.}.
In our following analysis we focus on the case where leptonic modes are dominant, choosing the small $v_{T_1}$ value since we are interested in the prediction for leptonic decay BRs in the model.
In addition, we assume the doubly charged scalar mass to be around TeV scale to avoid collider constraints~\cite{CMS:2017pet, Aaboud:2017qph, Ucchielli:2018koe}.
In this case the BRs for leptonic modes are simply given by~\cite{Padhan:2019jlc}
\begin{equation}
BR(\delta^{\pm \pm} \to \ell_i^\pm \ell_j^\pm) \simeq \frac{2}{1+\delta_{ij}} \frac{ |(m_\nu)_{ij}|^2}{\sum_{k,l } |(m_\nu)_{kl}|^2}, 
\end{equation}
where we ignored the decay width for $\delta^{\pm \pm} \to W^\pm W^\pm$ and $\delta_{ij}$ is the Kronecker delta.
We then estimate the BRs for the parameter sets which can accommodate the neutrino oscillation data in models (3) and (4).

\subsubsection{Model (3)}

\begin{figure}[tb]\begin{center}
\includegraphics[width=70mm]{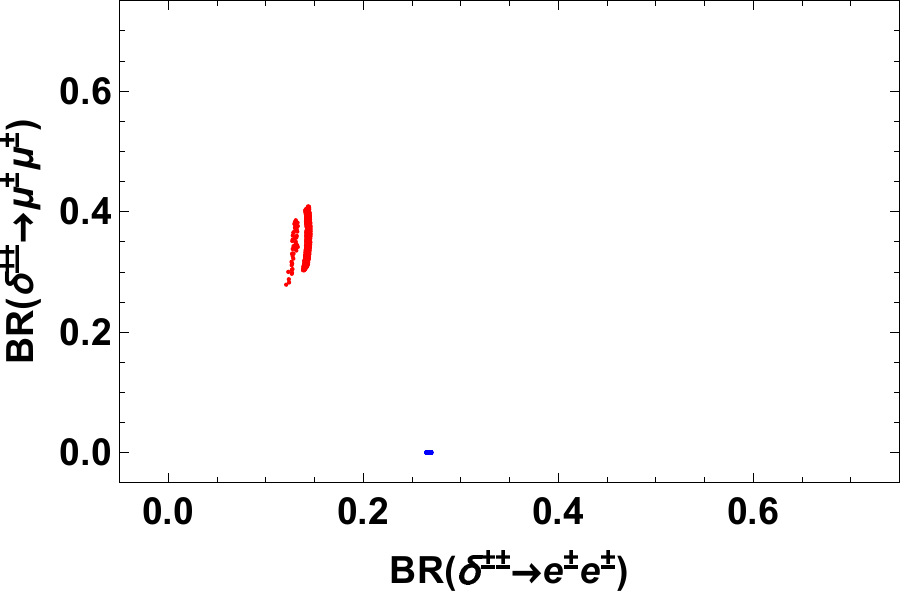} \
\includegraphics[width=70mm]{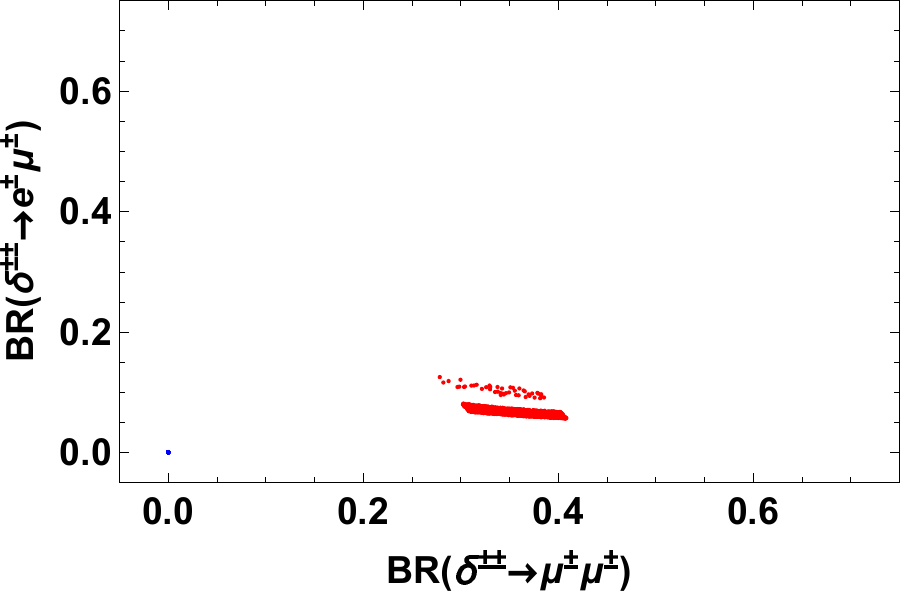} 
\includegraphics[width=70mm]{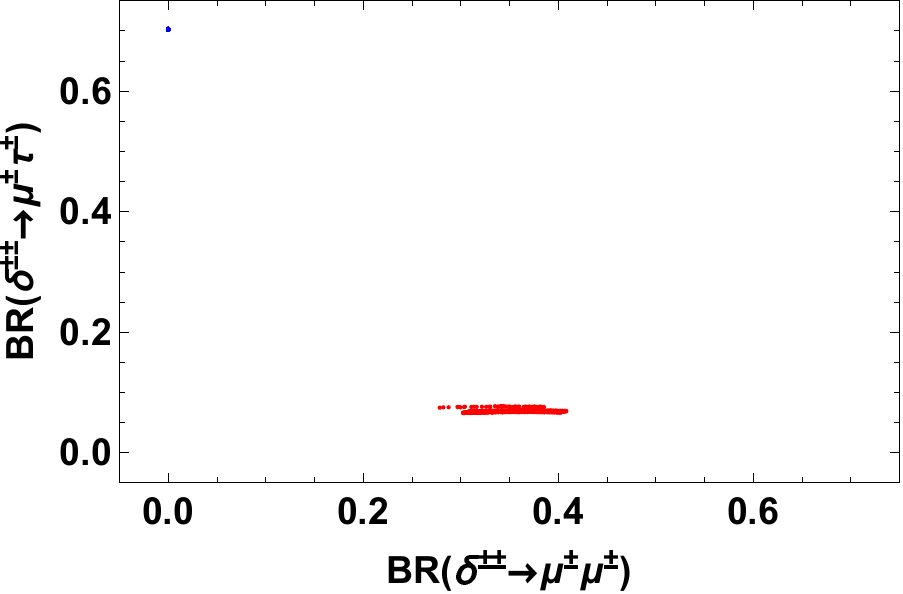} \
\includegraphics[width=70mm]{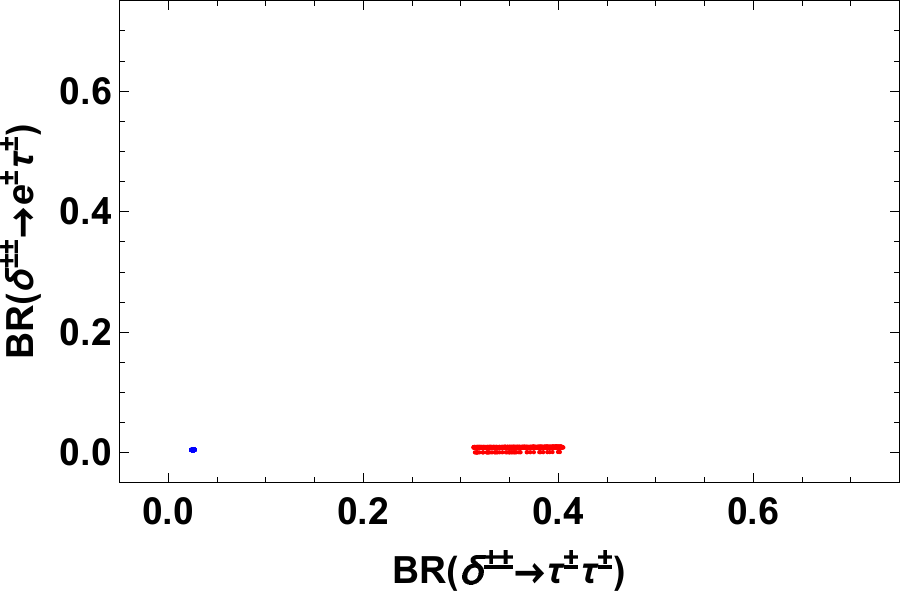} 
\caption{Predictions of doubly charged scalar BRs in model (3) case A for NO. The upper-left panel: predicted BRs for $\{ e^\pm e^\pm,  \mu^\pm \mu^\pm \}$ modes. 
The upper-right panel: predicted BRs for $\{ \mu^\pm \mu^\pm, e^\pm \mu^\pm \}$ modes. 
The lower-left panel: predicted BRs for $\{\mu^\pm \mu^\pm,  \mu^\pm \tau^\pm \}$ modes. 
The lower-right panel: predicted BRs for $\{ \tau^\pm \tau^\pm, e^\pm \tau^\pm \}$ modes.}   
\label{fig:BRM3NO}\end{center}\end{figure}

\begin{figure}[tb]\begin{center}
\includegraphics[width=70mm]{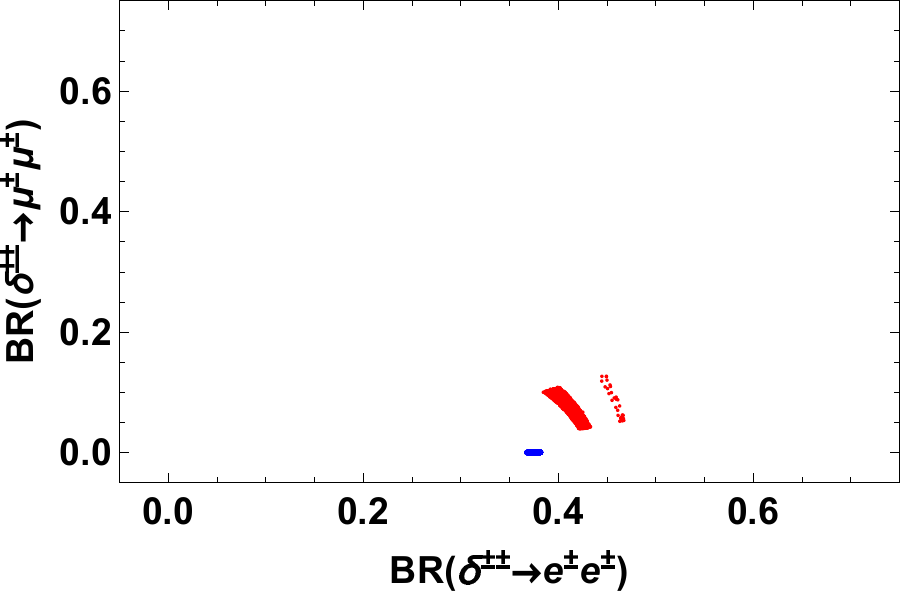} \
\includegraphics[width=70mm]{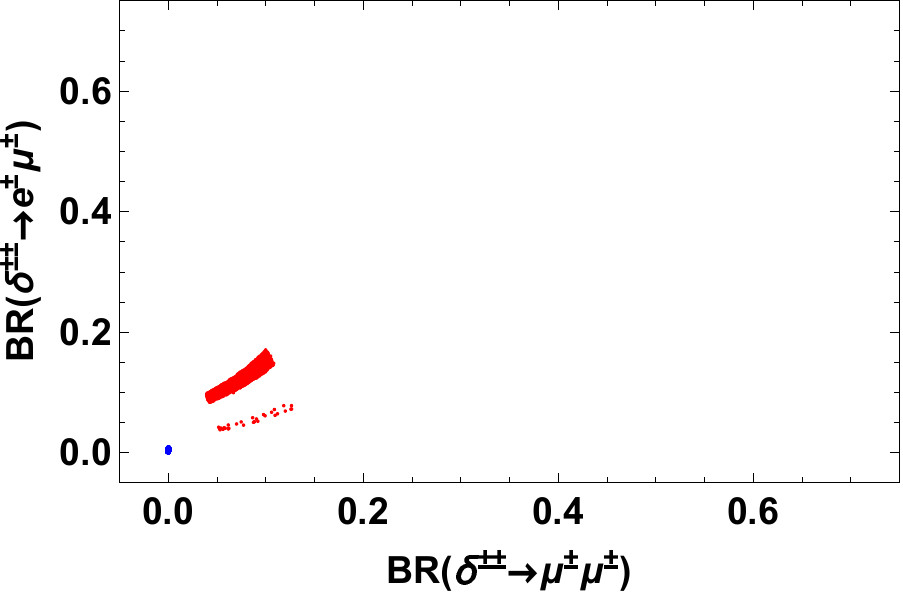} 
\includegraphics[width=70mm]{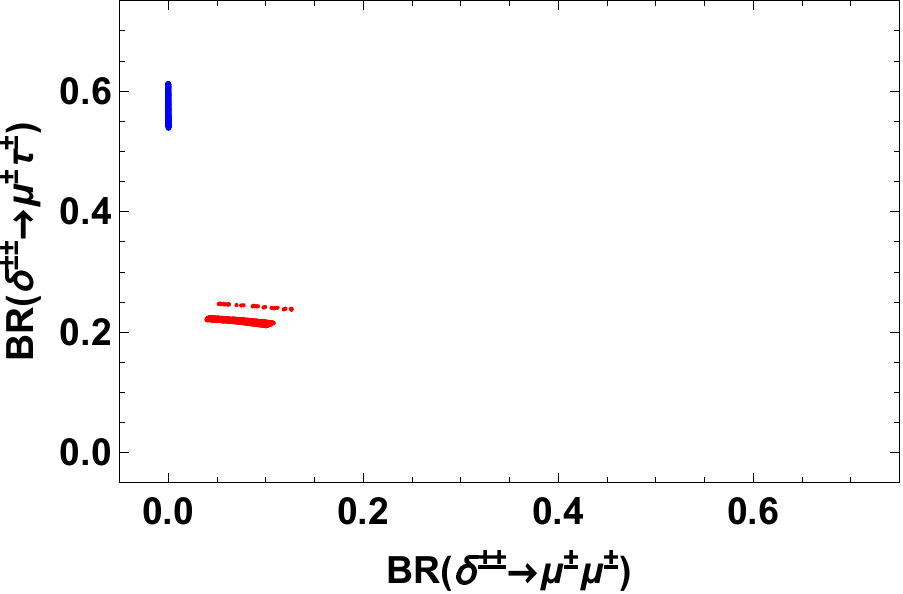} \
\includegraphics[width=70mm]{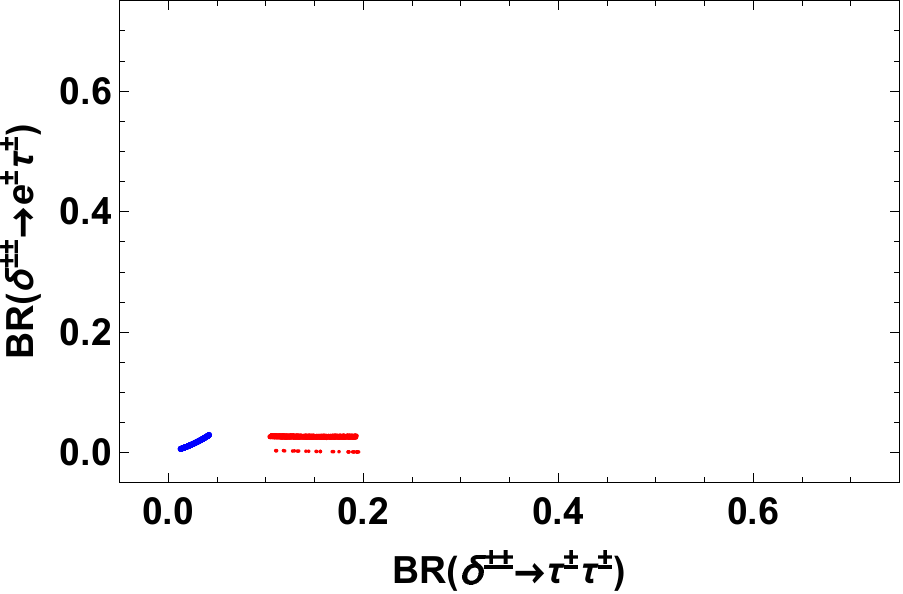} 
\caption{The same plots as Fig.~\ref{fig:BRM3NO} in the case of model (3) case A for IO.}   
\label{fig:BRM3IO}\end{center}\end{figure}

\begin{figure}[tb]\begin{center}
\includegraphics[width=70mm]{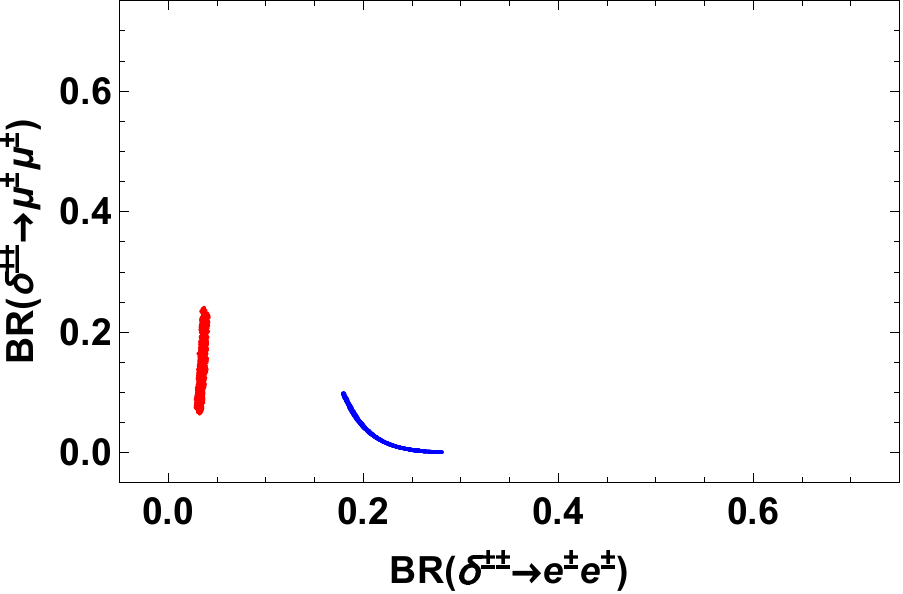} \
\includegraphics[width=70mm]{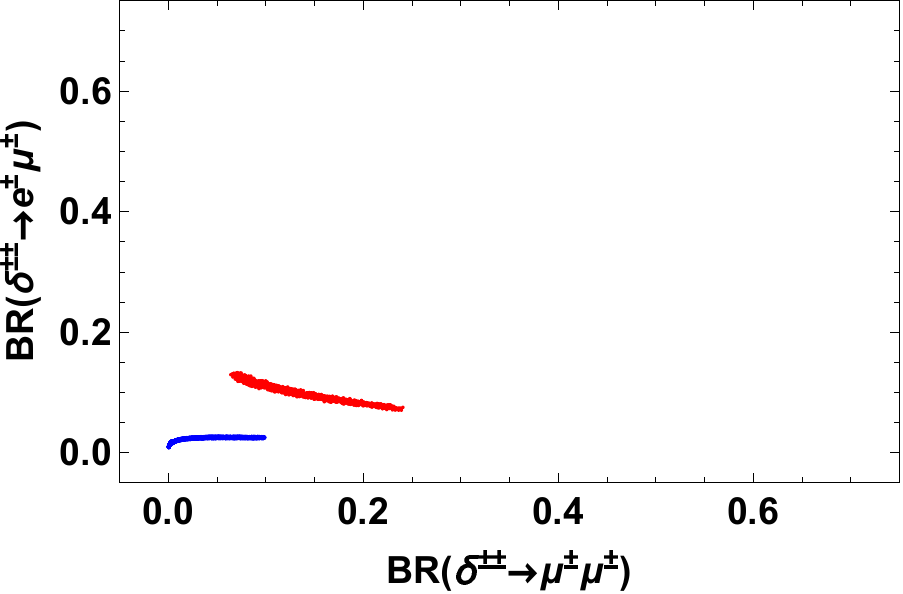} 
\includegraphics[width=70mm]{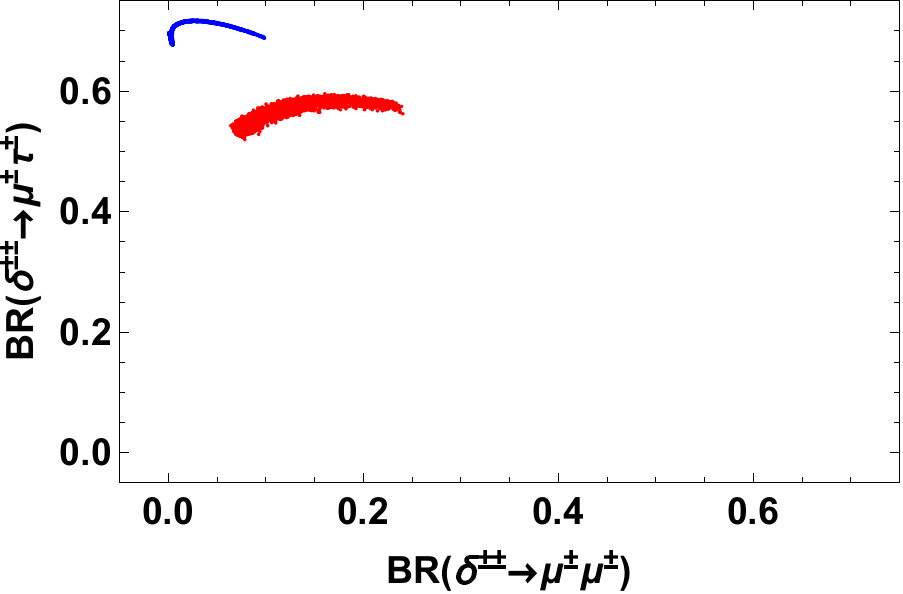} \
\includegraphics[width=70mm]{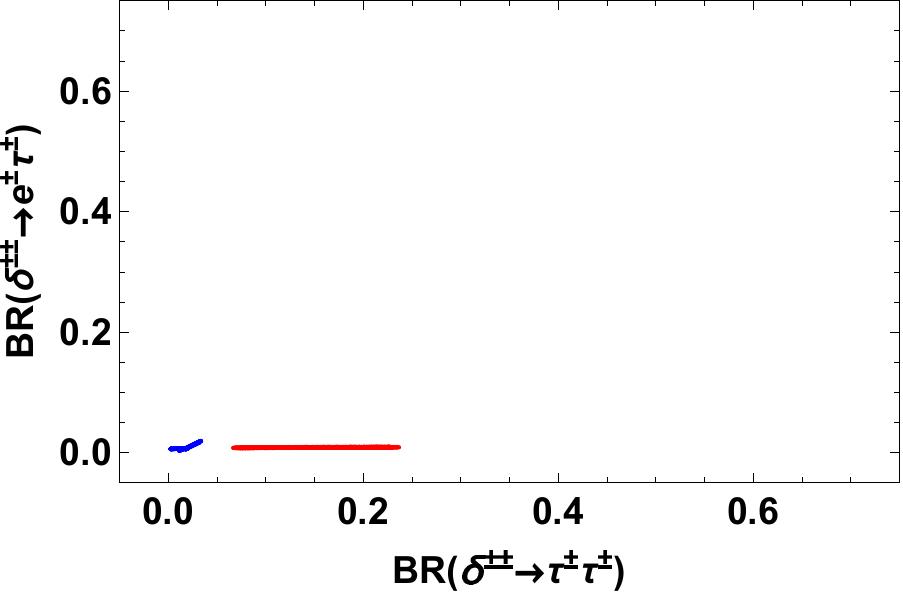} 
\caption{The same plots as Fig.~\ref{fig:BRM3NO} in the case of model (3) case B for NO.}   
\label{fig:BRM3NOB}\end{center}\end{figure}

\begin{figure}[tb]\begin{center}
\includegraphics[width=70mm]{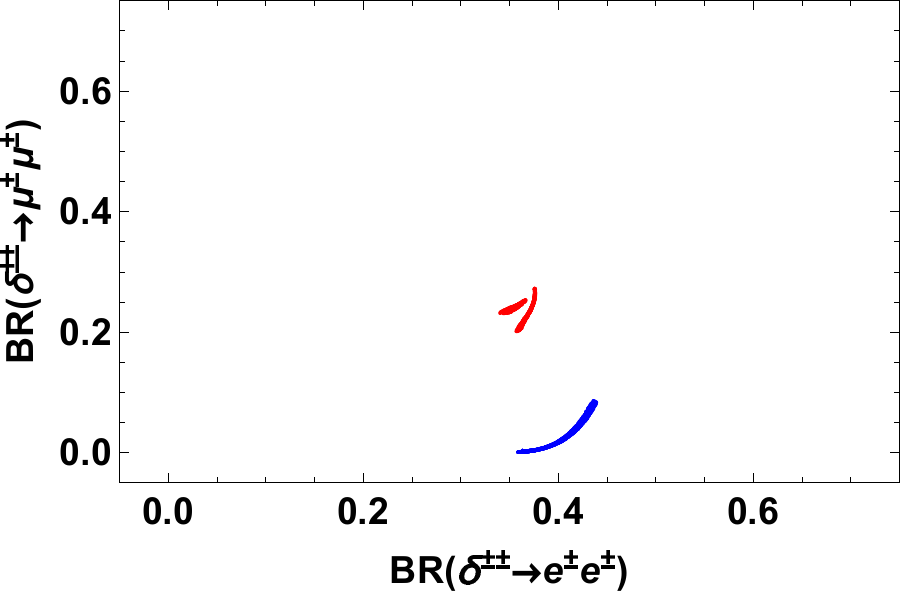} \
\includegraphics[width=70mm]{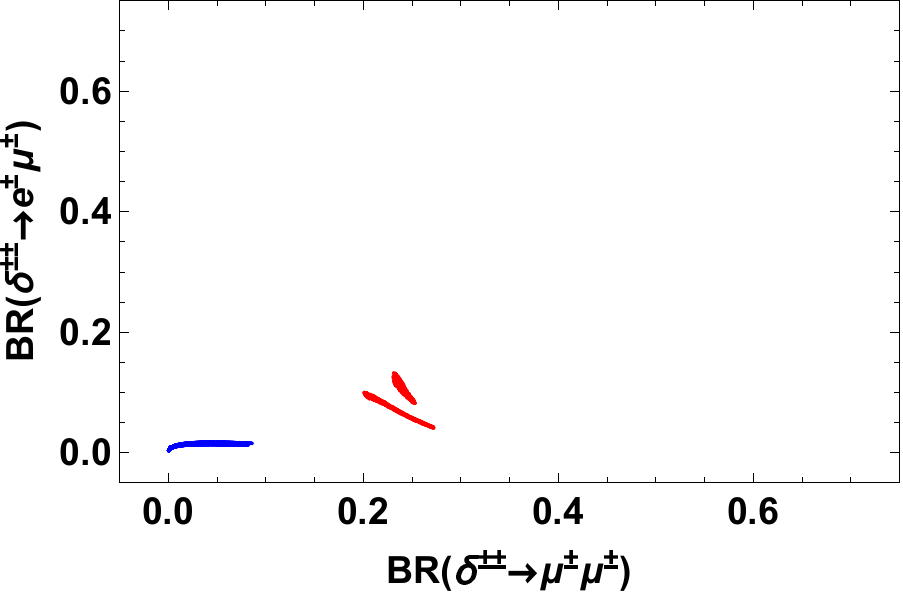} 
\includegraphics[width=70mm]{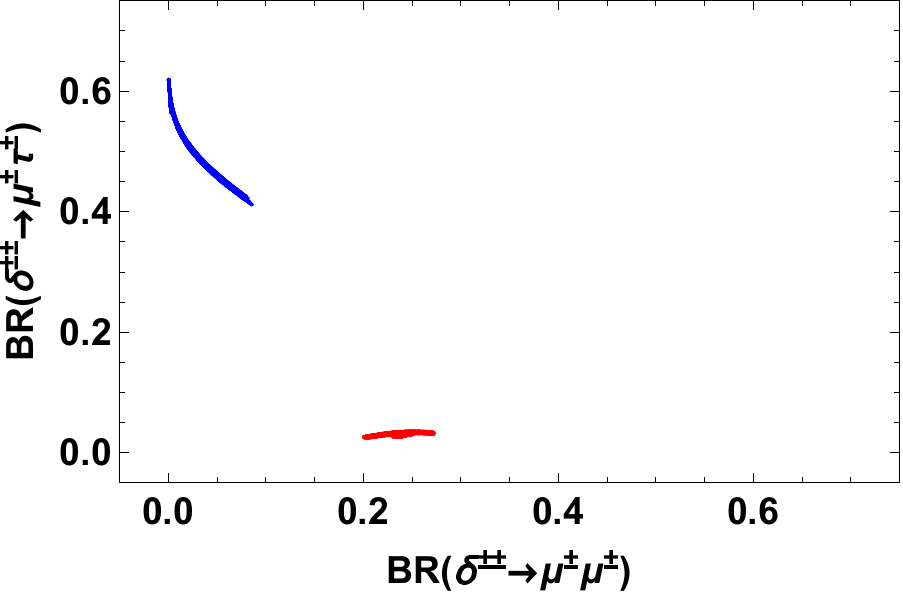} \
\includegraphics[width=70mm]{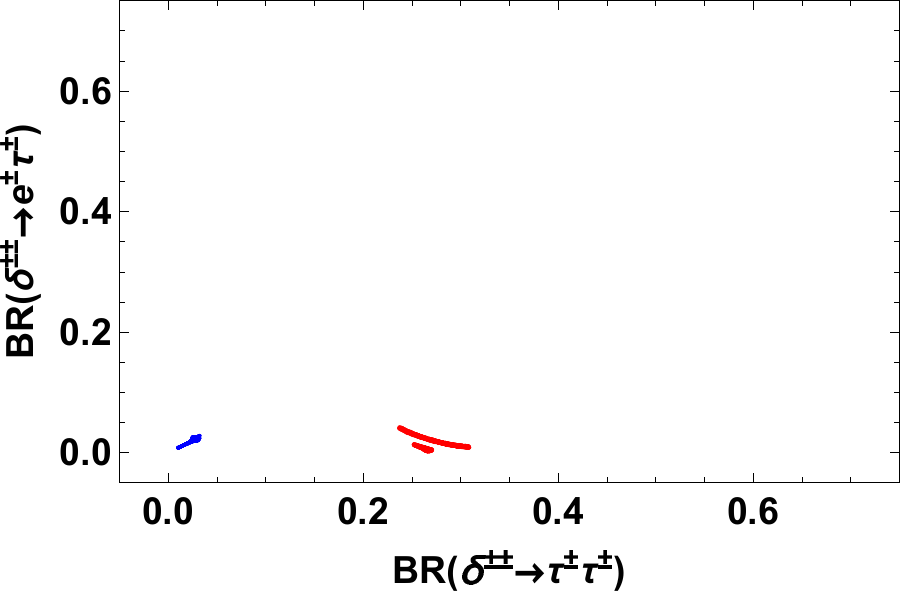} 
\caption{The same plots as Fig.~\ref{fig:BRM3NO} in the case of model (3) case B for IO.}   
\label{fig:BRM3IOB}\end{center}\end{figure}

In Fig.~\ref{fig:BRM3NO}, we show predicted BRs in case A for NO, applying the allowed parameter sets which accommodate the neutrino oscillation data.
In this case, we find two separate regions for the predicted BRs corresponding to two solutions of $\hat \alpha$ and $\hat \beta$ given by Eqs.~\eqref{eq:alphabeta1} and \eqref{eq:alphabeta2}. 
For Eq.~\eqref{eq:alphabeta1},  we obtain predicted region of BRs for modes $\{e^\pm e^\pm, \mu^\pm \mu^\pm, \tau^\pm \tau^\pm,  e^\pm \mu^\pm, e^\pm \tau^\pm, \mu^\pm \tau^\pm\}$ 
approximately as  $ \{0.1-0.15, 0.27-0.4, 0.3-0.4, 0.05-0.15, 0-0.02, 0.08-0.1 \}$; the order of BRs in the bracket will be the same in the following results.
For Eq.~\eqref{eq:alphabeta2},  we obtain the predicted region of BRs for modes approximately as $\{0.27, 0, 0.03, 0, 0, 0.7 \}$ which is more restricted.
In this case, the BR of the $e \tau$ mode is always small while values of the other modes depend on the solution of $\hat \alpha$ and $\hat \beta$. 

In Fig.~\ref{fig:BRM3IO}, we show predicted BRs in case A for IO applying the allowed parameter sets which accommodate the neutrino oscillation data.
In this case, we obtain the predicted region of the BRs approximately as $ \{0.39-0.47 0.05-0.13, 0.1-0.2, 0.04-0.2, 0-0.04, 0.2-0.25 \}$ for Eq.~\eqref{eq:alphabeta1} and 
 $ \{0.36-0.38, 0-0.01, 0.01-0.04, 0, 0-0.05, 0.54-0.61 \}$ for Eq.~\eqref{eq:alphabeta2}.
In this case, the BR of $e e$ and $\mu \tau$ modes are dominant processes. 

In Fig.~\ref{fig:BRM3NOB}, we show predicted BRs in case B for NO applying the allowed parameter sets.
In this case, we obtain predicted region of the BRs approximately as
approximately $ \{0.03-0.05, 0.07-0.24, 0.07-0.24, 0.1-0.15, 0, 0.52-0.6 \}$ for Eq.~\eqref{eq:alphabeta1} and $ \{0.18-0.28, 0-0.1, 0-0.03, 0-0.04, 0-0.03, 0.67-0.72 \}$ for Eq.~\eqref{eq:alphabeta2}.
In this case, BR of $\mu \tau$ mode is the dominant one for both solutions. 

In Fig.~\ref{fig:BRM3IOB}, we show the predicted BRs in case B for IO applying the allowed parameter sets.
In this case, we obtain predicted region of the BRs approximately as
 $ \{0.34-0.38, 0.2-0.27, 0.24-0.31, 0.05-0.15, 0-0.05, 0.04-0.05 \}$ for Eq.~\eqref{eq:alphabeta1} and $ \{0.35-0.44, 0-0.08, 0.01-0.03, 0-0.02, 0.01-0.03, 0.42-0.62 \}$ for Eq.~\eqref{eq:alphabeta2}.
In this case, the BR of the $e e$ mode is sizable for both solutions. BRs of flavor off-diagonal modes are not large  for Eq.~\eqref{eq:alphabeta1} while that of $\mu \tau$ mode is large for For Eq.~\eqref{eq:alphabeta2}. 

\subsubsection{Model (4)}

\begin{figure}[tb]\begin{center}
\includegraphics[width=70mm]{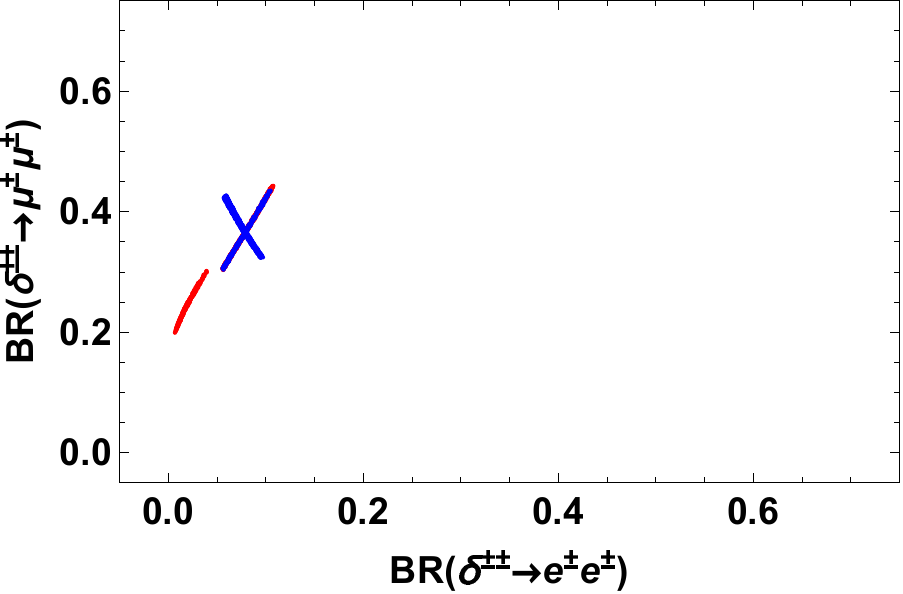} \
\includegraphics[width=70mm]{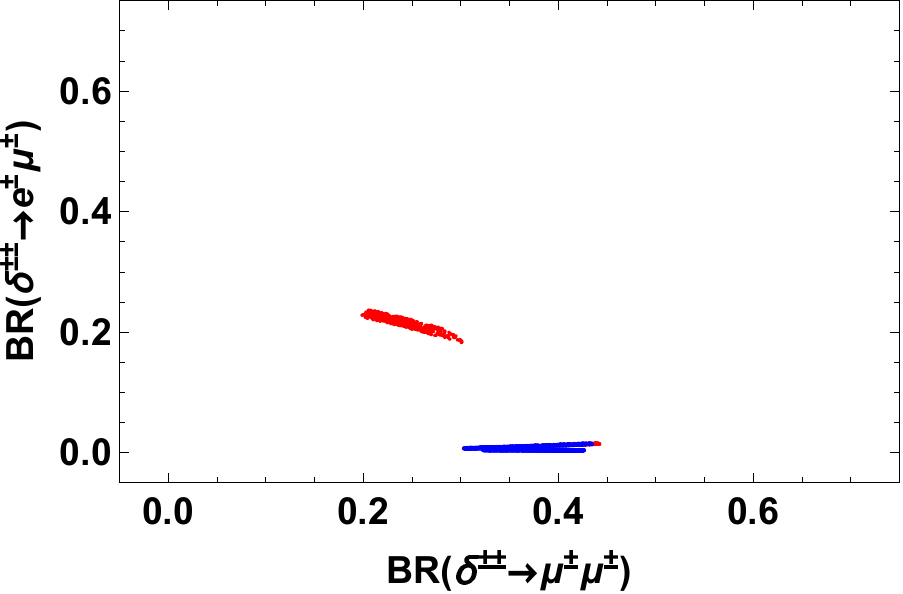} 
\includegraphics[width=70mm]{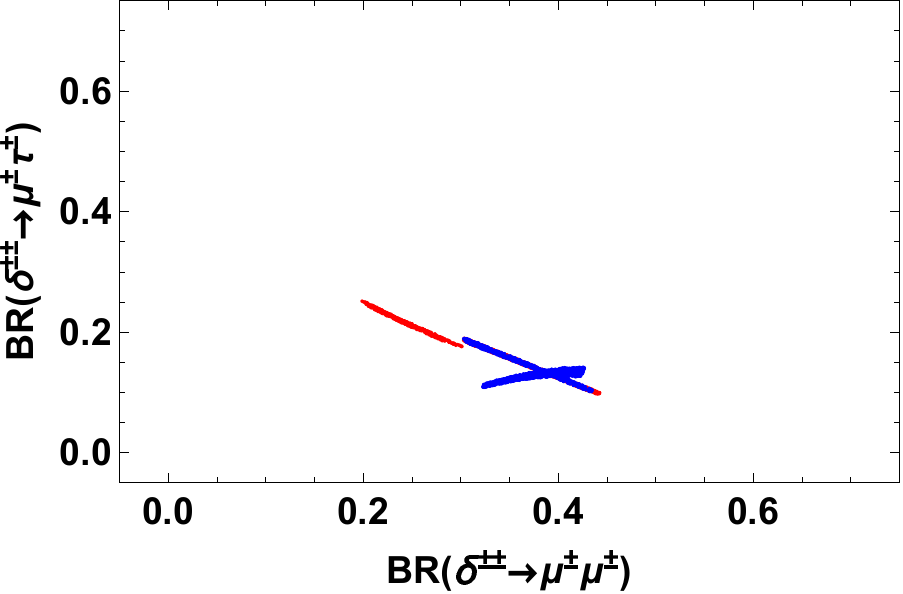} \
\includegraphics[width=70mm]{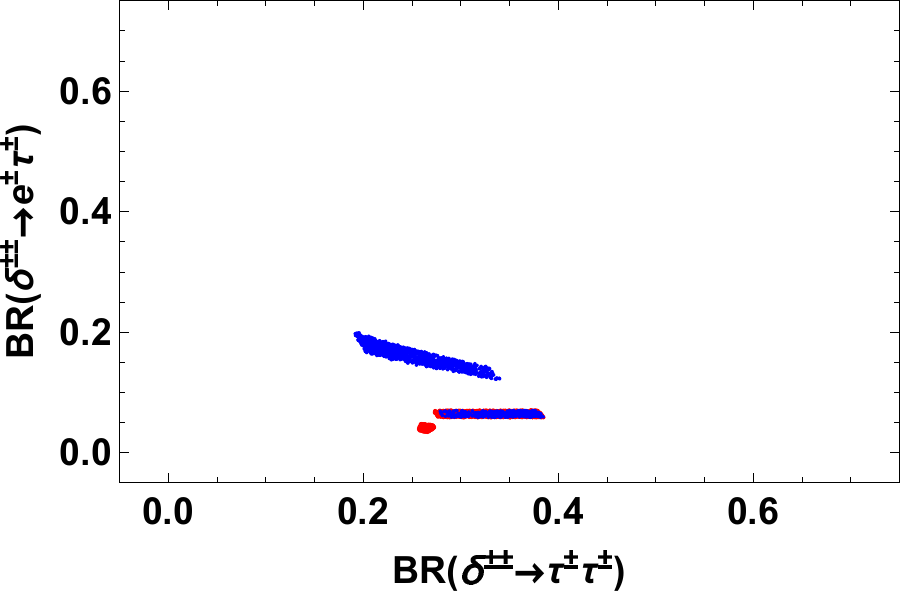} 
\caption{The same plots as Fig.~\ref{fig:BRM3NO} in the case of model (4) case A for NO.}   
\label{fig:BRM4NO}\end{center}\end{figure}

\begin{figure}[tb]\begin{center}
\includegraphics[width=70mm]{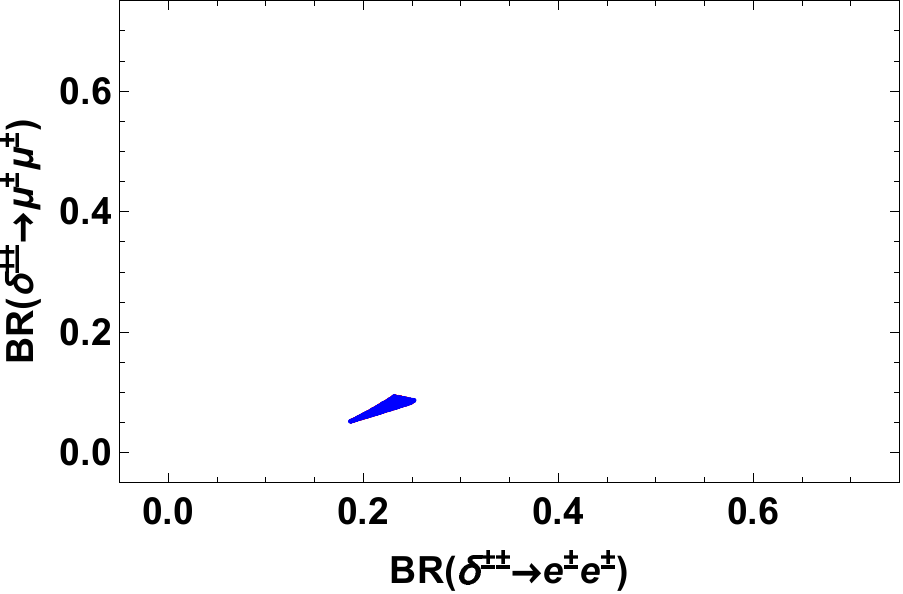} \
\includegraphics[width=70mm]{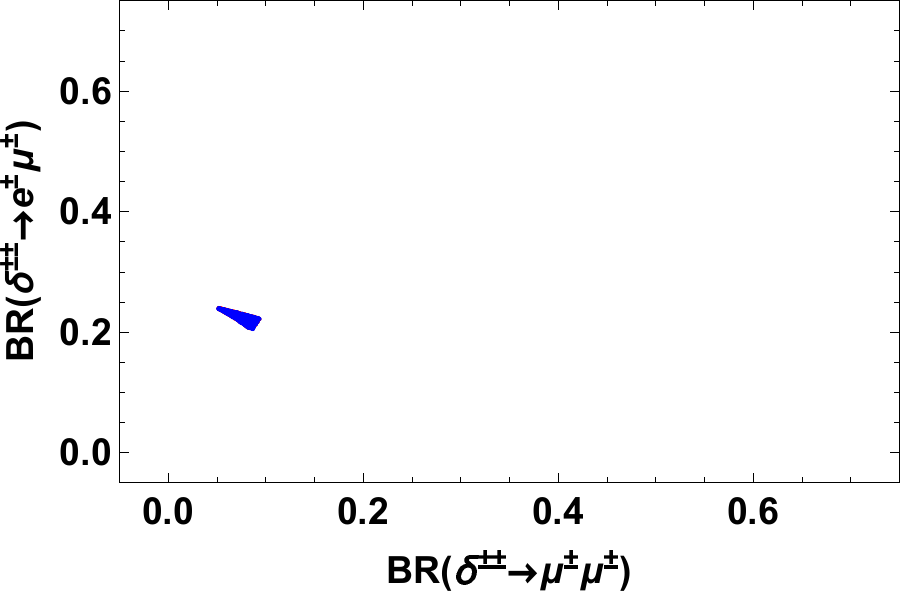} 
\includegraphics[width=70mm]{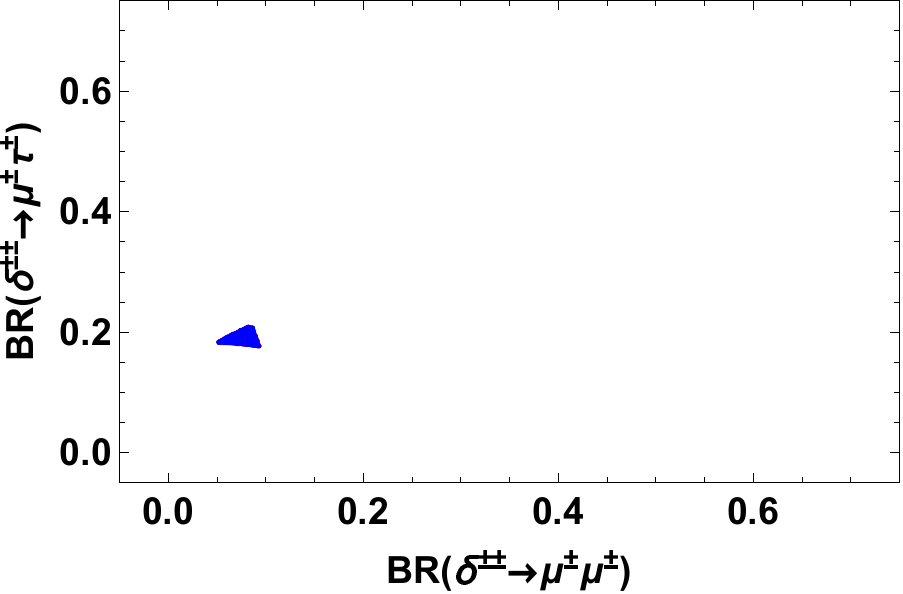} \
\includegraphics[width=70mm]{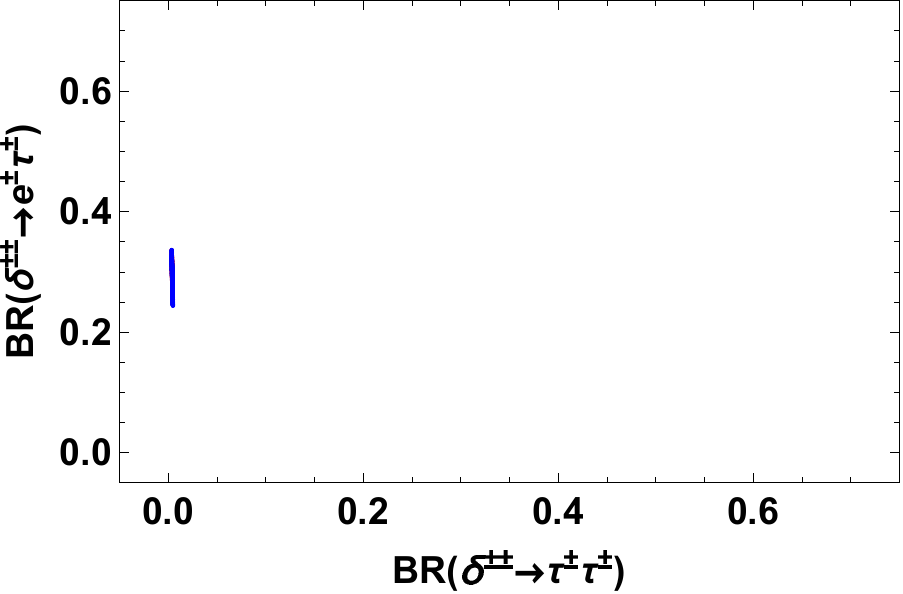} 
\caption{The same plots as Fig.~\ref{fig:BRM3NO} in the case of model (4) case A for IO.}   
\label{fig:BRM4IO}\end{center}\end{figure}

\begin{figure}[tb]\begin{center}
\includegraphics[width=70mm]{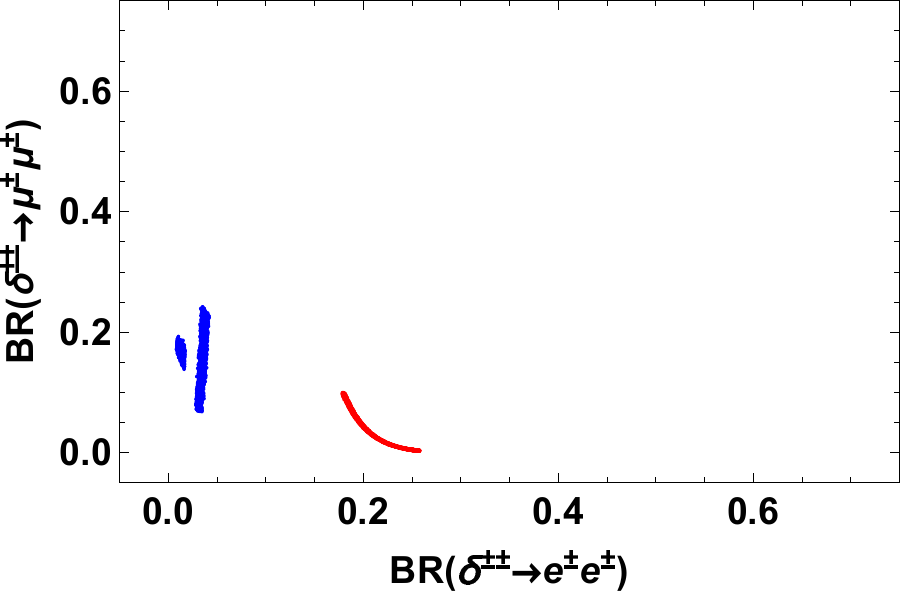} \
\includegraphics[width=70mm]{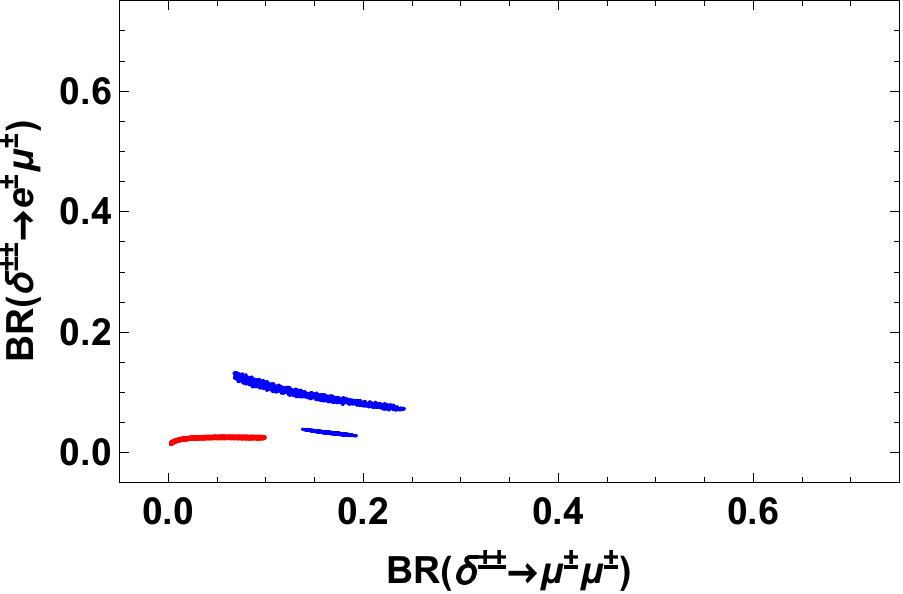} 
\includegraphics[width=70mm]{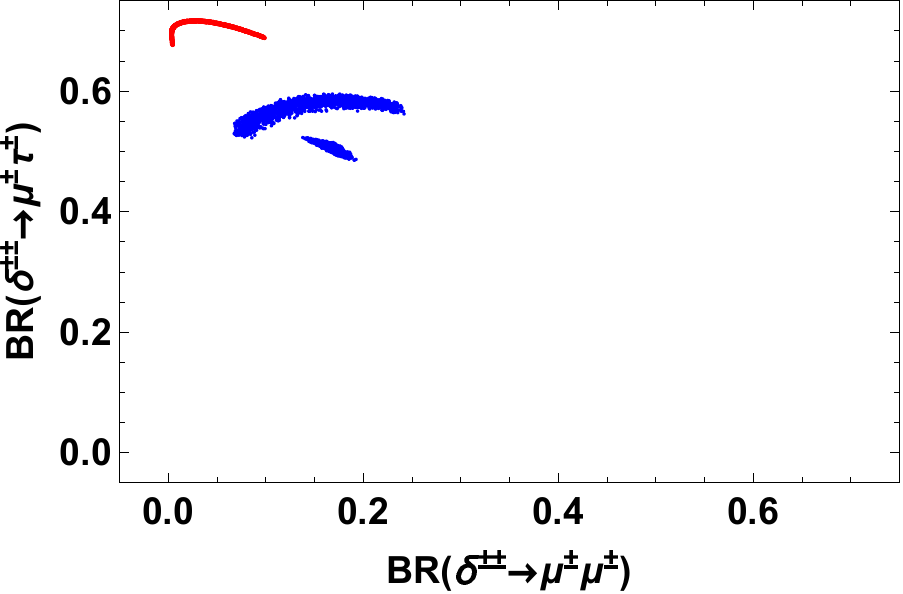} \
\includegraphics[width=70mm]{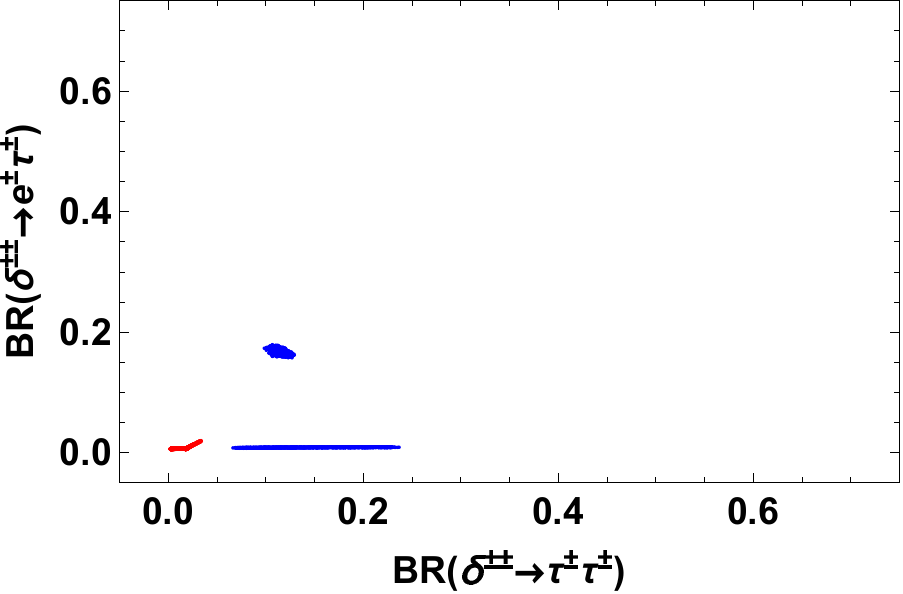} 
\caption{The same plots as Fig.~\ref{fig:BRM3NO} in the case of model (4) case B for NO.}   
\label{fig:BRM4NOB}\end{center}\end{figure}

\begin{figure}[tb]\begin{center}
\includegraphics[width=70mm]{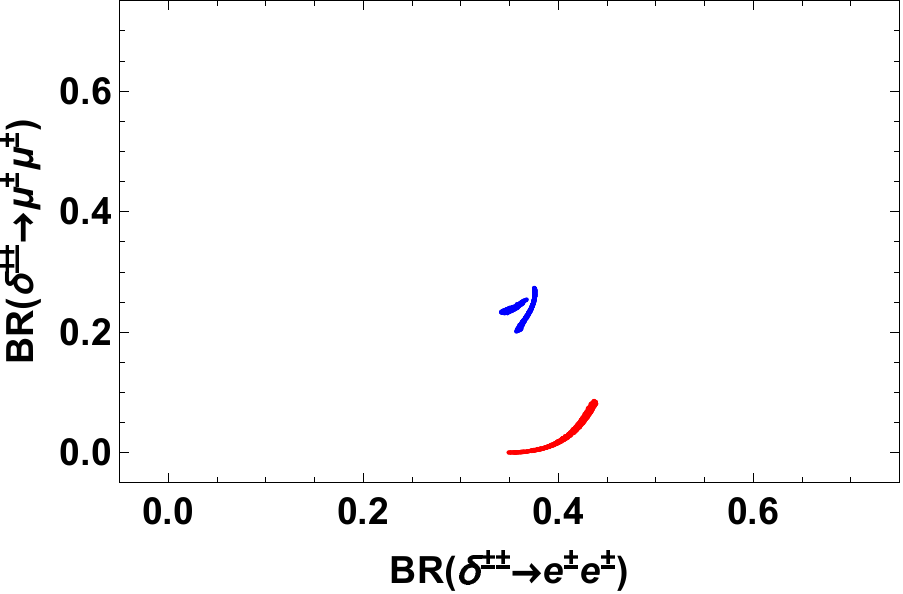} \
\includegraphics[width=70mm]{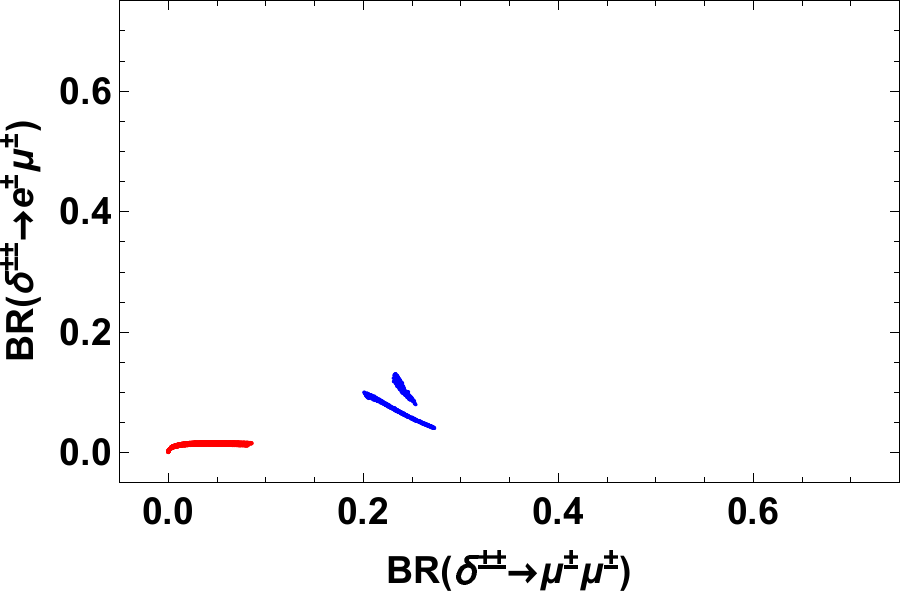} 
\includegraphics[width=70mm]{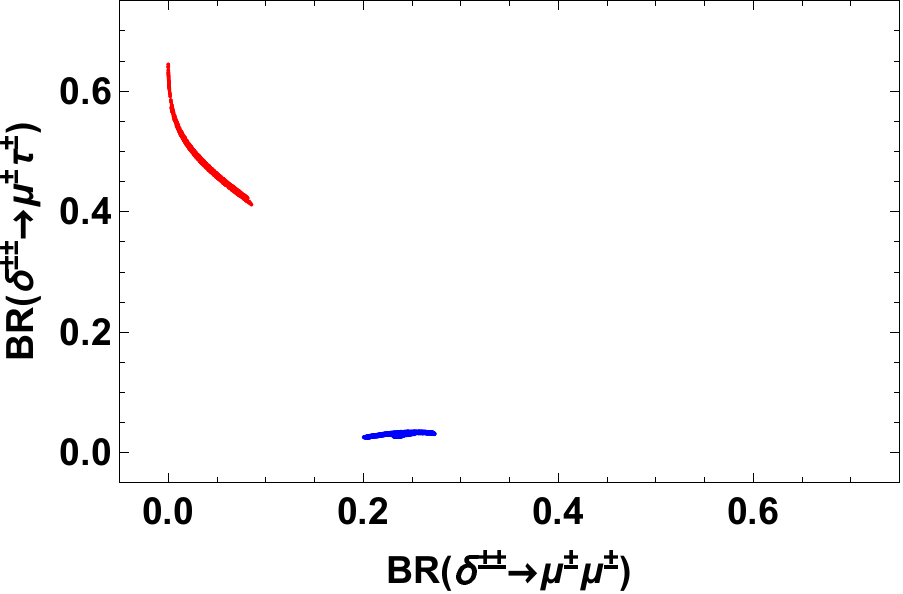} \
\includegraphics[width=70mm]{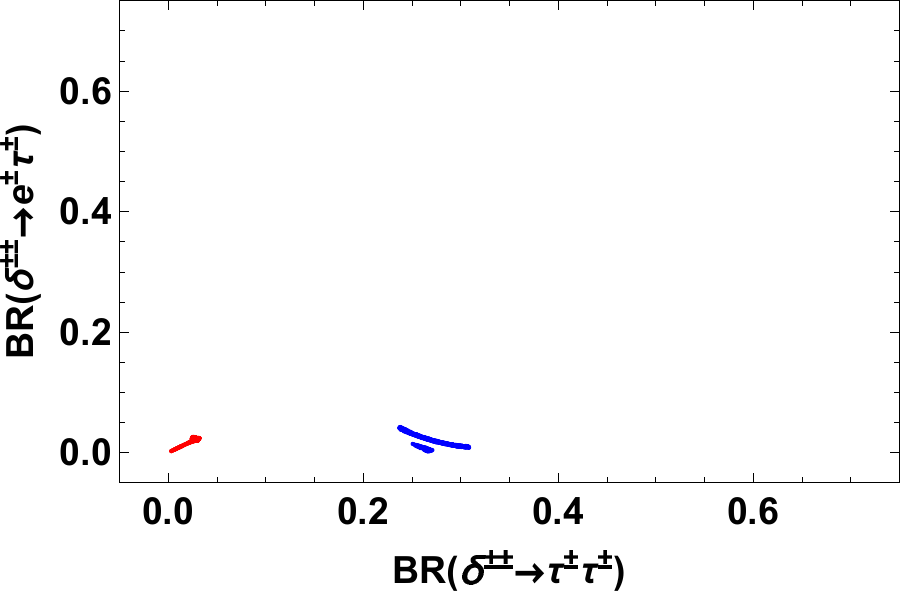} 
\caption{The same plots as Fig.~\ref{fig:BRM3NO} in the case of model (4) case B for IO.}   
\label{fig:BRM4IOB}\end{center}\end{figure}

In Fig.~\ref{fig:BRM4NO}, we show predicted BRs in case A for NO applying the allowed parameter sets.
In this case, we find several predicted regions of the BRs within an approximated range of
 $ \{0-0.1, 0.2-0.45, 0.25-0.38, 0-0.25, 0.03-0.07, 0.1-0.25 \}$ for Eq.~\eqref{eq:alphabeta1} and $ \{0.05-0.1, 0.3-0.44, 0.2-0.4, 0-0.03, 0.05-0.2, 0.1-0.2 \}$ for Eq.~\eqref{eq:alphabeta2}.
In this case, the BR of $\mu \mu$ and $\tau \tau$ modes are sizable for both solutions. For Eq.~\eqref{eq:alphabeta1}, $e \mu$ and $\mu \tau$ modes can be the dominant mode.

In Fig.~\ref{fig:BRM4IO}, we show the predicted BRs in case A for IO applying the allowed parameter sets.
In this case, we obtain the predicted region of the BRs approximately as
 $ \{0.2-0.25, 0.07-0.1, 0, 0.2-0.25, 0.25-0.34, 0.18-0.22 \}$ where the regions are the same for both Eq.~\eqref{eq:alphabeta1} and Eq.~\eqref{eq:alphabeta2} coincidentally.
In this case, the BR of the $\tau \tau$ modes is suppressed while the other modes can be sizable.

In Fig.~\ref{fig:BRM4NOB}, we show the predicted BRs in case B for NO applying the allowed parameter sets.
In this case, we obtain the predicted region of the BRs approximately as
approximately $ \{0.18-0.26, 0-0.1, 0-0.03, 0.01-0.04, 0-0.03, 0.67-0.72 \}$ for Eq.~\eqref{eq:alphabeta1} and $ \{0-0.05, 0.07-0.24, 0.07-0.24, 0.04-0.14, 0-0.18, 0.48-0.6 \}$ for Eq.~\eqref{eq:alphabeta2}.
In this case, the BR of the $\mu \tau$ mode is the dominant one for both solutions. 

In Fig.~\ref{fig:BRM4IOB}, we show predicted BRs in case B for IO applying the allowed parameter sets.
In this case, we obtain the predicted region of the BRs approximately as
$ \{0.35-0.44, 0-0.08, 0-0.03, 0-0.02, 0-0.03, 0.42-0.65 \}$ for Eq.~\eqref{eq:alphabeta1} and $ \{0.34-0.38, 0.2-0.27, 0.24-0.31, 0.05-0.14, 0-0.05, 0.03-0.05 \}$ for Eq.~\eqref{eq:alphabeta2}.
In this case, the BR of the $e e$ mode is sizable for both solutions. In addition, the $\mu \tau$ mode can be dominant for Eq.~\eqref{eq:alphabeta1} while flavor off-diagonal modes are suppressed for Eq.~\eqref{eq:alphabeta2}.

\section{Conclusion}
We have discussed type-II seesaw models with modular $A_4$ symmetry in the supersymmetric framework. 
In our approach, models are classified by the assignment of $A_4$ representations and modular weights for leptons and 
the triplet Higgs field.
Then, the free parameters in models are scanned to fit the neutrino oscillation data and we find the minimal cases including only the weight 2 modular form (model (1) and (2)) are disfavored.
We can fit the data for the models with weight 4 modular form (model (3) and (4)) applied to the neutrino mass matrix due to the additional two free parameters.
Then we have shown the predictions in the neutrino parameters for the allowed parameter sets accommodating the neutrino  oscillation data.
Finally, the branching ratios of the doubly charged scalar boson are calculated by applying the allowed parameter sets focusing on the case where the doubly charged scalar dominantly decays into charged leptons 
choosing the small triplet VEV.
We can predict the branching ratios where these values are realized to be in some restricted regions.
Therefore, it can be a clear indication of our models if we find the pattern of the branching ratios at the collider experiment. 
Furthermore we have the relations between predictions in the neutrino parameters and the branching ratios.
Importantly measurements of these branching ratios can test a flavor structure under the modular symmetry comparing predictions in the neutrino sector. 

Before closing the conclusion, we would like to comment on the other lepton flavor violating (LFV) processes 
such as $\mu \to e \gamma$, $\tau \to \mu(e) \gamma$ and $\mu \to 3 e$, etc. As shown in the branching ratios of 
the doubly charged scalar boson, the elements of the neutrino mass matrix are constrained and related with each other 
due to modular $A_4$ symmetry. Therefore it is expected that the branching ratios of the LFV processes will also have correlations 
among them. If it is the case, such correlations will provide more useful information which enables us to discriminate 
our model from others. The LFV decay branching ratios depend on the mass of the charged scalar bosons and hence 
need detailed analyses 
including the spectrum of the scalar bosons.
Such analyses are beyond the scope of this paper and we will leave this for our future work.
We also would like to comment on potential corrections to our predictions. When modular $A_4$ symmetry 
is broken simultaneously with supersymmetry breaking, the supersymmetry breaking mechanism affects 
the predictions on the neutrino oscillation observables. One of such corrections will come from threshold corrections to 
the charged lepton mass in large $\tan\beta$ case. In addition, renormalization group evolution of the parameters also affects the predictions. 
The RGE effects depend on a full spectrum of particles and can be important when the supersymmetric particles are relatively light. 
When we take into account these effects, our results may change. However, these analyses are also beyond the scope 
of this paper, and will be studied elsewhere.

\appendix

\section{Multiplication rule of $A_4$ group} \label{apdx:multiplication-rule}
In this appendix, we give generators and multiplication rules of modular $A_4$ symmetry used in our calculation.
Throughout this paper, we employ the three dimensional unitary representation in the so-called $T$-diagonal basis as 
\begin{align}
T = 
\begin{pmatrix}
1 & 0 & 0 \\
0 & \omega & 0 \\
0 & 0 & \omega^2 
\end{pmatrix}, ~~~~
S = \frac{1}{3}
\begin{pmatrix}
-1 & 2 & 2 \\
2 & -1 & 2 \\
2 & 2 & -1
\end{pmatrix}.
\end{align}

The multiplication rule for  a product of $A_4$ triplet representations in this basis is given by,
\begin{align}
\begin{pmatrix}
a_1 \\
a_2 \\
a_3
\end{pmatrix}_{\bf{3}}
\otimes 
\begin{pmatrix}
b_1 \\
b_2 \\
b_3
\end{pmatrix}_{\bf{3}}
&=
(a_1 b_1 + a_1 b_3 + a_3 b_2 )_{\bf{1}} 
\oplus (a_3 b_3 + a_1 b_2 + a_2 b_1 )_{\bf{1'}} \nonumber \\
&\quad \oplus (a_2 b_2 + a_1 b_3 + a_3 b_1 )_{\bf{1''}} \nonumber \\
&\quad \oplus 
\frac{1}{3}
\begin{pmatrix}
2 a_1 b_1 - a_2 b_3 - a_3 b_2 \\
2 a_3 b_3 - a_1 b_2 - a_2 b_1 \\
2 a_2 b_2 - a_1 b_3 - a_3 b_1 
\end{pmatrix}_{\bf{3}}
\oplus \frac{1}{2}
\begin{pmatrix}
a_2 b_3 - a_3 b_2 \\
a_1 b_2 - a_2 b_1 \\
a_3 b_1 - a_1 b_3 
\end{pmatrix}_{\bf{3}},
\end{align}
and those for products of singlet representations are given  by
\begin{align}
\bf{1} \otimes \bf{1} = \bf{1},~~~\bf{1'} \otimes \bf{1'} = \bf{1''},~~~\bf{1''} \otimes \bf{1''} = \bf{1'},
~~~\bf{1'} \otimes \bf{1''} = \bf{1}.
\end{align}
More details can be found in \cite{Ishimori:2010au,Ishimori:2012zz}.

\section{Determining free parameters in charged lepton mass matrix} \label{apdx:lepton-mass}

In this appendix we summarize the determination of free parameters, $\{ \alpha, \beta, \gamma\}$, in 
the charged lepton mass matrix in Eq.~(\ref{eq:ME1}) following discussion in ref.~\cite{Kobayashi:2018scp}.
We have three equations with the charged lepton mass eigenvalues:
\begin{subequations}
\begin{align}
& \text{Tr}[M_E^\dagger M_E] = \sum_{i=e}^{\tau} m_i^2 = \frac{| \tilde \gamma Y_3|^2}{4} (1+ \hat \alpha^2 + \hat \beta^2) C_1, \label{eq:cond1} \\
& \text{Det}[M_E^\dagger M_E] = \prod_{i=e}^{\tau} m_i^2 = \frac{|\tilde \gamma Y_3|^6}{64} \hat \alpha^2 \hat \beta^2 C_2, \\
& \frac{\text{Tr}[M_E^\dagger M_E]^2 - \text{Tr}[M_E^\dagger M_E]}{2} = \chi = \frac{|\tilde \gamma Y_3|^4}{16} (\hat \alpha^2 + \hat \alpha^2 \hat \beta^2 + \hat \beta^2) C_3, 
\end{align}
\label{eq:charged-lepton-mass-invariants}
\end{subequations}
where $\chi \equiv m_e^2 m_\mu^2 + m_\mu^2 m_\tau^2 + m_\tau^2 m_e^2$.
The coefficients $C_1$, $C_2$ and $C_3$ are given by $\hat Y_2 \equiv Y e^{i \phi_Y}$, where $Y$ is real positive and $\phi_Y$ is a phase parameter, such that
\begin{subequations}
\begin{align}
& C_1 = (2+Y^2)^2, \\
& C_2 = 64 + 400 Y^6 + Y^{12} - 40 Y^3 (Y^6 - 8) \cos (3 \phi_Y) - 16 Y^6 \cos (6 \phi_Y), \\
& C_3 = 16 + 16 Y^2 + 36 Y^4 + 4 Y^6 + Y^8 - 8 Y^3 (Y^2 - 2) \cos (3 \phi_Y).
\end{align}
\label{eq:charged-lepton-coefficients}
\end{subequations}
The values of these coefficients are determined when we fix the value of modulus $\tau$. 
We then obtain the general equations to determine $\hat \alpha$ and $\hat \beta$:
\begin{equation}
\frac{(1+s)(s+t)}{t} = \frac{(\sum m_i^2/C_1) (\chi/C_3)}{\prod m^2_i/C_2}, \quad \frac{(1+s)^2}{s+t} = \frac{(\sum m^2_i/C_1)^2}{\chi/C_3},
\end{equation}
where $s \equiv \hat \alpha^2 + \hat \beta^2$ and $t \equiv \hat \alpha^2 \hat \beta^2$.
We thus obtain $\hat \alpha$ and $\hat \beta$ by the relation:
\begin{align}
\label{eq:alphabeta1}
\hat \alpha^2_1 = \frac{s + \sqrt{s^2 - 4t}}{2}, \quad \hat \beta^2_1 = \frac{s - \sqrt{s^2 - 4t}}{2}, \\
\label{eq:alphabeta2}
\hat \alpha^2_2 = \frac{s - \sqrt{s^2 - 4t}}{2}, \quad \hat \beta^2_2 = \frac{s + \sqrt{s^2 - 4t}}{2}, 
\end{align}
where we separately write the possible two solutions for $\hat \alpha$ and $\hat{\beta}$.
Finally $\tilde \gamma$ is determined by $\hat \alpha$ and $\hat \beta$ via Eq.~(\ref{eq:cond1}).

For the charged lepton Yukawa couplings with modular weight $k=4$, Eqs.~\eqref{eq:charged-lepton-mass-invariants} 
and \eqref{eq:charged-lepton-coefficients} are given as follows,
\begin{subequations}
\begin{align}
& \text{Tr}[M_E^\dagger M_E] = \sum_{i=e}^{\tau} m_i^2 = \frac{| \tilde \gamma Y_3^2|^2}{4} (1+ \hat \alpha^2 + \hat \beta^2) C_1, \label{eq:cond2} \\
& \text{Det}[M_E^\dagger M_E] = \prod_{i=e}^{\tau} m_i^2 = \frac{|\tilde \gamma Y_3^2|^6}{64} \hat \alpha^2 \hat \beta^2 C_2, \\
& \frac{\text{Tr}[M_E^\dagger M_E]^2 - \text{Tr}[M_E^\dagger M_E]}{2} = \chi = \frac{|\tilde \gamma Y_3^2|^4}{16} (\hat \alpha^2 + \hat \alpha^2 \hat \beta^2 + \hat \beta^2) C_3, 
\end{align}
\label{eq:charged-lepton-mass-invariants2}
\end{subequations}
where
\begin{subequations}
\begin{align}
& C_1 = \frac{1}{4} (Y^8 + 4 Y^6 + 36 Y^4 + 16 Y^2 - 8 (Y^2 - 2) Y^3 \cos(3 \phi_Y) + 16), \\
& C_2 = \frac{1}{64} (Y^{12} - 16 Y^6 \cos(6 \phi_Y) + 400 Y^6 - 40 (Y^6-8) Y^3 \cos(3 \phi_Y) + 64)^2, \\
& C_3 = \frac{1}{16} (Y^2 + 2)^2 (Y^{12} - 16 Y^6 \cos(6 \phi_Y) + 400 Y^6 - 40 (Y^6-8) Y^3 \cos(3 \phi_Y) + 64 ).
\end{align}
\label{eq:charged-lepton-coefficients2}
\end{subequations}
Similarly $\tilde \gamma$ is determined by $\hat \alpha$ and $\hat \beta$ via Eq.~(\ref{eq:cond2}).

\section*{Acknowledgments}
\vspace{0.5cm}

This work is supported by 
MEXT KAKENHI Grant Numbers JP19H04605 (T.K.), 
JP18H05543 (T.S.), and 
JSPS KAKENHI Grant Nos.~JP18H01210 (T.S.),  
JP18K03651 (T.S.).



\end{document}